\documentclass[
 reprint,
 amsmath,amssymb,
 aps,
 floatfix
]{revtex4-1}

\usepackage{graphicx}
\usepackage{dcolumn}
\usepackage{bm}
\usepackage{url}
\usepackage{comment}
\usepackage{here}

\begin{document}

\preprint{APS/123-QED}

\title{Short-term memory by transient oscillatory dynamics in recurrent neural networks}

\author{Kohei Ichikawa}
 
\author{Kunihiko Kaneko}
 \altaffiliation{kaneko@complex.c.u-tokyo.ac.jp; also, at Research Center for Complex Systems Biology, Universal Biology Institute, University of Tokyo.} 
\affiliation{
 Graduate School of Arts and Sciences, The University of Tokyo, 3-8-1 Komaba, Meguro-ku, Tokyo 153-8902, Japan.
}

\date{\today}

\begin{abstract}
Despite the significance of short-term memory in cognitive function, the process of encoding and sustaining the input information in neural activity dynamics remains elusive. Herein, we unveiled the significance of transient neural dynamics to short-term memory. By training recurrent neural networks to short-term memory tasks and analyzing the dynamics, the characteristics of the short-term memory mechanism were obtained in which the input information was encoded in the amplitude of transient oscillations, rather than the stationary neural activities. This transient trajectory was attracted to a slow manifold, which permitted the discarding of irrelevant information. Additionally, we investigated the process by which the dynamics acquire robustness to noise. In this transient oscillation, the robustness to noise was obtained by a strong contraction of the neural states after perturbation onto the manifold. This mechanism works for several neural network models and tasks, which implies its relevance to neural information processing in general.

\end{abstract}

\maketitle

\section{Introduction}
Short-term memory is essential for our cognitive activities\cite{review-of-short-term-memory}. 
Once an external signal is applied to an internal neuron, its
information is encoded therein and saved for a certain time period.
Some form of sustained neural activity is therefore
necessary\cite{doi:10.1152/jn.1996.76.5.2841}. 
A conventional theory of working memory is based on the representation of each memory item as a different attractor of the considered system\cite{Funahashi1989, Fuster1971, fixed-point-short-term-memory}. 
This theory is usually referred to as ``memories as attractors.’’  
However, this is not practical when a large number of objects has to be memorized, because it is difficult to prepare so many attractors.
Additionally, if there is a need to memorize continuous information (e.g., length, frequency, amplitude of inputs), it becomes more difficult to form a memory using this attractor.
Furthermore, extensive experimental reports suggest that while short-term memory is maintained, neural activities are not constant; they instead continue to change over time\cite{HAIDER2009171, Dynamics_of_Population_Code}. 
Indeed, the possibility of achieving short-term memory through sustained transient dynamics has recently been discussed\cite{ostojic2020, Druckmann2012, Goldman2009, Orhan2020Improved, Orhan2019}. 
Additionally, the general relevance of transient activities to neural information processing has been discussed\cite{Rabinovich48}.

It is unclear, however, what form of transient neural activity can afford short-term memory. 
In contrast to established studies on memories as attractors, the way external inputs are encoded into transient dynamics is not well explored. Furthermore, in contrast to stationary neural activities at attractors, the way information is sustained in the transient process with time-varying neural activities remains elusive. 
Memory must be robust under noise to inputs or internal neural activities. From the view of memories as attractors, the stability of memory is supported as the state is attracted to the attractors (i.e., stationary states) after perturbation\cite{barak_fixed_point, Sussillo2019}, whereas the robustness in the transient dynamics is not well-explored. One must investigate how robustness to noise is achieved to understand the transient dynamics that support short-term memory.

To investigate the neural dynamics of short-term memory, we adopted a recurrent neural network (RNN) trained to solve a task that requires memorizing the input information for a given time span. We provided a task to compare two interspaced signals input with some time interval. The RNN is required to determine which of the two subsequent signals has the larger continuous characteristics (e.g., frequencies of the periodic signals)\cite{romo}. Here, to solve the comparison task, it is necessary for the neural dynamics to maintain the information of the first signal until the second signal input arrives. Thus, short-term memory is needed to solve this comparison task. In this RNN, the neural dynamics consist of activities entailing a large number of neurons connected via synapses whose weights are adjusted by the standard back-propagation method\cite{Rumelhart:1986we, bptt} to solve this task. 

After the network is trained to solve the task successfully, we analyze the generated neural dynamics to uncover how the input information is encoded and memorized according to the dynamics of neural activities. 
We find that the neural activities exhibit a transient oscillation that endures for the time span between the first and second signals. 
Subsequently, we analyze how this oscillatory neural activity encodes the input information, provides short-term memory, and solves the task. We uncover that the information of the first signal (e.g., its frequency) is encoded and memorized by the amplitude of the transient oscillation.
Then, we determine if and how this memory by the amplitude of transient oscillation is robust to noise using dynamical systems theory.

The remainder of this paper is organized as follows. 
In the next section, we introduce the task to compare the frequencies of two signals input with a given time interval as well as the RNN model used to solve it. 
After demonstrating that the trained RNN can solve the task, we analyze how the memory of the first input is maintained. 
We demonstrate that neural activities during the time span, while memory is maintained, exhibit a transient oscillation. Here, the input-signal information required to solve the task is encoded by the amplitude of this transient oscillation, whereas the irrelevant information in the input signal that is unrelated to solving the task is discarded. Then, the robustness of the memory to noise is analyzed. 
We also confirm that the mechanism for short-term memory presented in this study is valid for several different comparison tasks as well as in neural network models that take biological features into the account, such as excitatory-inhibitory balance or sparseness. 
Finally, the possible relevance of the presented mechanism for short-term memory to biological neural dynamics is discussed.

\section{model}
\subsection{Short-term memory task}
\begin{figure}[b]
    \includegraphics[width=8cm]{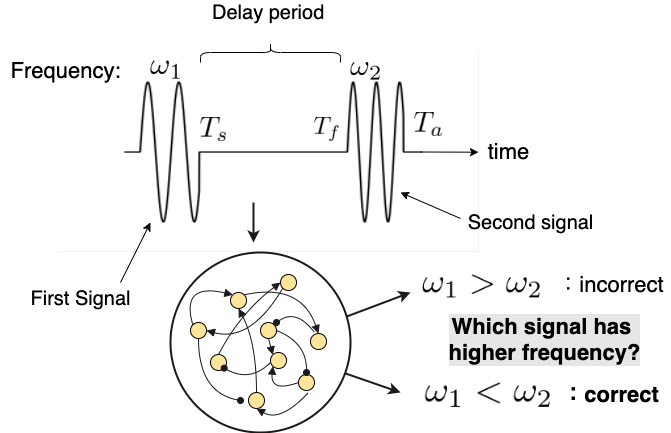}
\caption{
         \label{fig:fig1} Schematic diagram of frequency comparison task. The input consists of the first signal, a delay period during which there is no input signal, and the second signal. Both the first and second signals are represented by noisy sine waves. The task is to determine which of the frequencies of the first and second signals is higher. The frequencies of two signals satisfy $1\leq \omega_1, \omega_2 \leq 5$. In addition, only in the training phase the difference of two signals satisfy $\ |\omega_1-\omega_2| \geq 1$. 
        }

\end{figure}

As a task that requires short-term memory, we studied the frequency comparison task illustrated in Fig.\ref{fig:fig1}, which is commonly adopted in the field of neuroscience\cite{romo}. 
In this task, the input signals consist of the first signal, a delay period, and the second signal. The objective of the task is to determine which frequency is higher: that of the first or the second signal.

Specifically, the first and second signals are chosen as noisy sine waves following $u_{1,2}(t)=\sin(\omega_{1,2} t + \phi)+\eta_{1,2}(t)$, where $\phi_{1,2}$ represents the phase of the signals, and $\eta_{1,2}(t)$ is a random Gaussian variable with average zero and standard deviation 0.05. 
During the delay period, there is no input signal. 
The duration of the first(second) signal, $T_s(T_s')$ and the delay period, $T_d$, vary with each sample in the training phase. 
At $T_s$, the delay period starts, and at $T_f=T_s+T_d$, the delay period ends. 
At $T_a=T_s+T_d+T_s'$, the second signal completes, and the neural network is asked whether the first or second has the higher frequency. 
Specifically, $T_s$ is homogeneously distributed as $T_s \in [13,17]$, and $T_d$ is homogeneously distributed as $T_d \in [25,35]$ throughout the samples in the training phase(see \ref{table:table1}). 
During the test phase, to answer the task, $T_s$ and $T_d$ are fixed at $15$ and $30$, respectively. 
These conditions are fixed unless otherwise mentioned.

\begin{table}
    \begin{center}
        \begin{tabular}{|c|c|c|} \hline
            condition &  training phase & test phase \\ \hline
            Length of the input signal &   $13\leq T_s \leq 17$ & $T_s=15$ \\ \hline
            Length of the delay period &   $25\leq T_d \leq 35$ & $T_d=30$ \\ \hline
            Frequencies of signals &   $1\leq \omega_1, \omega_2 \leq 5$ & $1\leq \omega_1, \omega_2 \leq 5$ \\ \hline
            Difference of $\omega_1,\omega_2$ &  $\ |\omega_1-\omega_2| \geq 1$ & $\ |\omega_1-\omega_2| \geq 0$ \\ \hline
        \end{tabular}
        \caption{
            \label{table:table1} Conditions of the short-term memory task.
        }
    \end{center}
\end{table}
\subsection{Recurrent Neural Network}
In this study, a standard model\cite{PhysRevLett.118.258101} is adopted for neural activity dynamics as expressed in the equation

\begin{equation}
    \dot{x}_i = -x_i + \sum_{j=1}^N J_{ij}\tanh(x_j)+W_i^{\rm in}u,
\end{equation}

where $x_i$ represents the neural activity state of neuron $i$ (or $(1+\tanh x_i)/2$ can be correlated with the firing rate)\footnote{Because $x_i$ takes both positive and negative values, it does not represent the neural activity per se. 
Instead, in the context of neuroscience, $(1+\tanh(x_i))/2$ is often regarded as the firing rate of the neuron.}. 
Each neuron is recurrently connected to the others, where $J_{ij}$ represents the strength of the connection from neuron $j$ to $i$. 
Furthermore, $u_i$ represents the input signal, which is defined by a cognitive task as described above and is projected onto the neurons by the input connection, $W^{\rm in}_i$. The number of neurons, $N$, is set at 256.
The initial state of neuron $x_i(t=0; i=1,...,N)$ is set to be a Gaussian random variable with average zero and standard deviation of 0.1.
In this study, the Euler method is applied to simulate Eq. (1), and the discretized dynamics are subsequently calculated. A discretized Eq.(1) is adopted here as $x_i(t+1)  = (1-\alpha)x_i(t)+ \alpha (\sum_{j=1}^N J_{ij}\tanh(x_j(t))+W_i^{in}u(t)) $. The time width, $\alpha$, for discretization is set to 0.25; nonetheless, the choice of this specific value is not essential to the results.

The output of the RNN is determined by the weighted sum of the neural states, as stated in the equation,
\begin{equation}
    z_i(t) = \sum_{j=1}^N W_{ij}^{out}x_j.
\end{equation}

$\bf z$ is chosen to be a two-component vector, and $z_1-z_2$ represents the logit of the probability that the first signal has a higher frequency\cite{PRML}. 
The calculation from ${\bf z}$ to the probability is performed using the softmax function, ${\rm Softmax}(z_i)=e^{z_i}/(e^{z_1}+e^{z_2})$\footnote{The relationship between logit $z_1-z_2$ and the probability calculated by the softmax function is as follows. By definition of the softmax function, we have $p_1=e^{z_1}/(e^{z_1}+e^{z_2})=1/(1+e^{-(z_1-z_2)})$. 
Then, by the definition of logit function, we get the logit of $p_1$ as $\log(p_1/(1-p_1))=\log(e^{z_1-z_2})=z_1-z_2$. 
From the above explanation, the output of the softmax function can be treated as a probability because $p_1$ and $p_2$ satisfy $0\leq p_1,p_2 \leq 1$, and $p_1+p_2=e^{z_1}/(e^{z_1}+e^{z_2})+e^{z_2}/(e^{z_1}+e^{z_2})=1$. In this study, specifically, we interpret this probability as the probability that the first signal is recognized to have a higher frequency by the RNN.}. 
Therefore, ${\rm Softmax}(z_i)$ represents the probability that the RNN judge that the $i$-th signal has the higher frequency;
If $z_1>z_2$, the RNN estimates the first signal to have a higher frequency, and vice versa.

The parameters of this RNN are composed of the weights of the vector, ${\bf W}^{\rm in}$, and two matrices, ${\bf J}$ and ${\bf W}^{\rm out}$; these weights are adjusted by training the RNN to solve the task introduced in the previous subsection. The specific learning procedure is described in the following subsection. Although, in the reservoir computing scheme\cite{echo-state, FORCE_learning} training only changes the output matrix, ${\bf W}^{\rm out}$, in this study, we adjust the weights of all three matrices.

\subsection{Training}
To train the RNN for the short-term memory task described above, we adopt a stochastic gradient descent scheme, which is commonly used in machine-learning communities\cite{DeepLearning}. In this scheme, the loss function is first defined to indicate how far the output of the current RNN is from the correct answer. This loss function is given as a function of the weights of the matrices, which provide the parameters of the RNN. The learning process is carried out by calculating the gradient and optimizing the parameters in the direction toward which the loss function becomes smaller. Although there are many optimization algorithms, the basic concept is given by the following equation:

\begin{equation}
    W_{t+1} = W_t - \eta \nabla_w {\mathcal L},
\end{equation}
where $W$ represents the parameters of the RNN (i.e., vector ${\bf W}^{\rm in}$ and two matrices ${\bf J}$ and ${\bf W}^{\rm out}$), and $\eta$ represents the learning rate. $\nabla_w {\mathcal L}$ represents the gradient of the loss function and is calculated via back-propagation through time (BPTT) \cite{bptt}, in which the dynamics of the RNN are first unfolded in time; then, the derivatives of the loss function are calculated using the chain-rule. In this study, we adopt the softmax cross-entropy loss function:
\begin{equation}
{\mathcal L}_{\rm CE} = -\sum_{k=1}^2 \hat{z}_k \log z_k^{\rm softmax},
\end{equation}
where $z_k^{\rm softmax} \equiv e^{z_k}/(e^{z_1} + e^{z_2})$, and $\bf \hat z$ represents the target label of this task. If $\omega_1$ is larger than $\omega_2$, $\hat{\bf z} = (1, 0)^{\mathrm T}$; otherwise, $\hat{\bf z} = (0, 1)^{\mathrm T}$. This target label is defined to satisfy the demand of this task. As explained, $z_k^{\rm softmax}$ gives the probability that the $k$-th signal is recognized to have higher frequency by the RNN. The softmax cross-entropy loss, therefore, represents the difference between the probability estimated by the RNN and the target probability. This loss function is widely used owing to its computational efficiency\cite{deep_learning_book}.

To make the training process stable, L2 norm regularization\cite{PRML} is applied for the sum of the norm of the RNN weights. Finally, the loss function is defined as  

\begin{equation}
     {\mathcal L} = {\mathcal L}_{\rm CE} + \lambda_{\rm L2}(\sum_i W_{{\rm in}, i}^2+\sum_{ij}J_{ij}^2+\sum_{ij}W_{{\rm out}, ij}^2),
\end{equation}
where $\lambda_{\rm L2}$ is set to 0.0001. During the training phase, the loss function is summed up for 50 input samples; then, the weights of the matrices are adjusted by the gradient descent. This process is continued for 3,000 iterations. Here we adopt ADAM\cite{adam} as the specific algorithm of optimization, which uses $\nabla_w {\mathcal L}$ calculated by BPTT, as is mentioned above. The learning rate, $\eta$, is set to 0.001. The training algorithm is implemented using PyTorch\cite{NEURIPS2019_9015}, which is the framework for machine learning.

\begin{figure}
    \includegraphics[width=7cm]{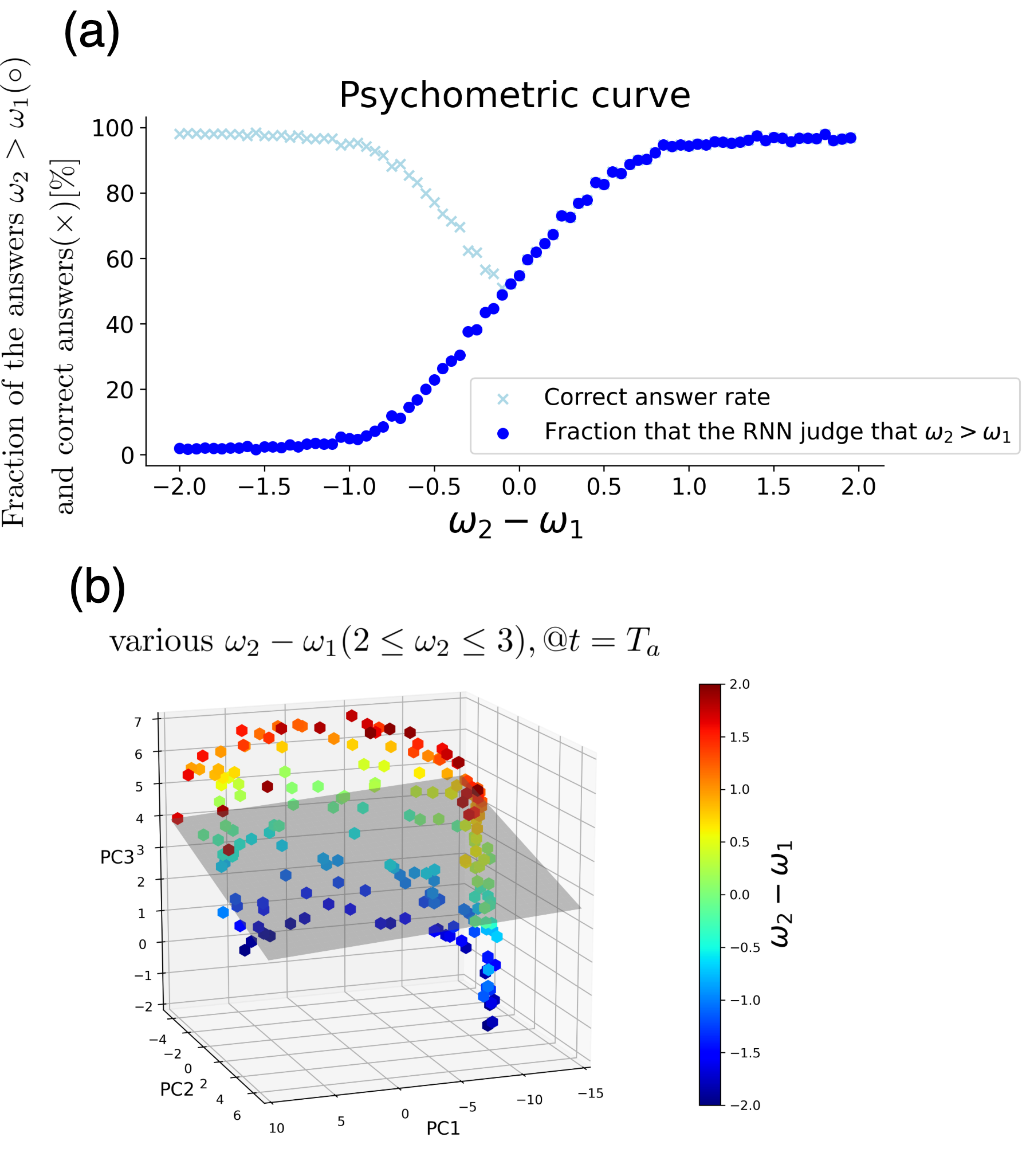}
    \caption{
        \label{fig:fig2} (a) Fraction judged by the RNN judge to be $\omega_2 \geq \omega_1$($\circ$). The horizontal axis represents the difference between $\omega_1$ and $\omega_2$. Hence, if the fraction is 100\% (0\%) for $\omega_2-\omega_1>0(<0)$, respectively, the correct answer rate ($\times$) is 100\%. As shown, if $|\omega_1-\omega_2|\geq 1 $, the correct answer rate is almost $100\%$, whereas it decreases as $|\omega_1-\omega_2|$ decreases. (b) Neural states $\bf x$ at the end of the second signal ($t=T_a$; timing to judge), plotted in the 3-dimensional principal component (PC) space. The PC axis is computed from the data of the neural states at $T_a$ for various frequency inputs. Colors represent $\omega_2-\omega_1$. The plane in the center can separate these neural states into those satisfying $\omega_1>\omega_2$.
        }

\end{figure}

\section{Results}
\subsection{Short-term memory by transient oscillatory dynamics}
After training, 50 pairs of signals having various $\omega_1$ and $\omega_2$ were input to the trained RNN. 
In Fig.\ref{fig:fig2}a, the fraction judged by the RNN to be $\omega_2$ was higher than $\omega_1$ plotted against $\omega_2-\omega_1$. 
With the condition of $|\omega_2-\omega_1|>1$, the accuracy of the choice was greater than 95\%. 
Hence we can see that the RNN correctly learned to solve the frequency comparison task. 
To more closely examine how the neural dynamics compared the frequencies, the neural states corresponding to various $\omega_1$ and $\omega_2$ at $t=T_a$ (i.e., at the end of the second signal) were plotted using the three principal components of $\{x_i\}$(Fig.\ref{fig:fig2}b)\cite{PCA}. The state changed continuously in response to the difference in frequencies. Moreover, according to $\omega_1> \omega_2$ or $\omega_1 \leq \omega_2$, the states could be separated by a plane. Hence, the frequency of the first signal was memorized over the delay period.

\begin{figure*}
    \includegraphics[width=16.5cm]{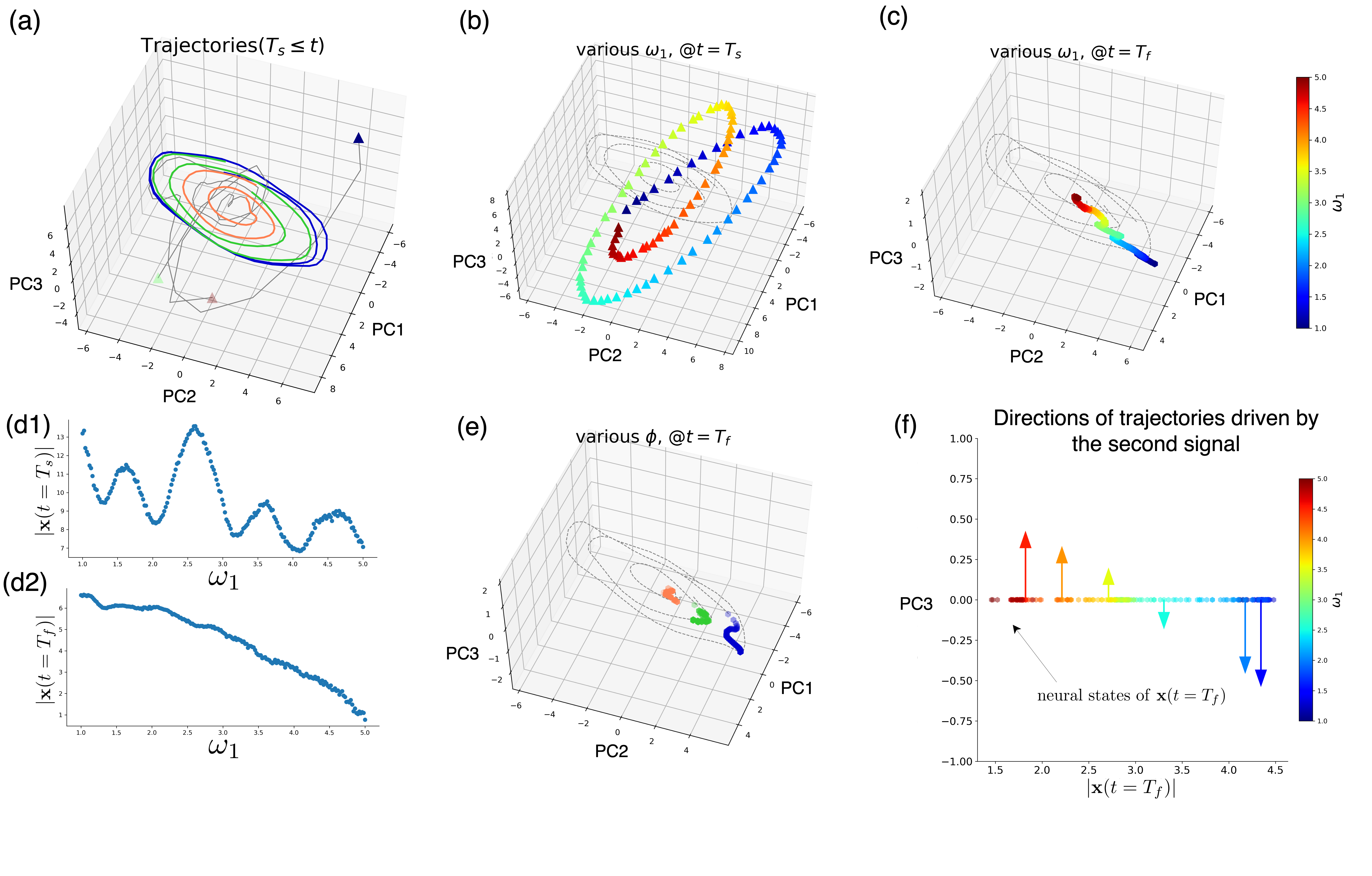}
    \caption{
        \label{fig:fig3} (a) Trajectories of neural activities during a prolonged delay period ($T_s \leq t$, without second signal) were plotted in a 3-dimensional principal component (PC) space. Trajectories from three different $\omega_1$, giving rise to different states at $t=T_s$, are given by triangle symbol. (b) Neural states $\bf x$ in the PC space at the beginning of the delay period ($t=T_s$) for 50 different $\omega_1$'s presented with different colors. (c) Neural states $\bf x$ in the PC space at the end of the delay period ($t=T_f$) for $\omega_1$ corresponding to (b). (d1) Scatter plot of the norm, $|{\bf x}(t=T_s)|$, against the frequency of the first signal, $\omega_1$. (d2) Scatter plot of the norm at the end of the delay period, $|{\bf x}(t=T_f)|$, against the frequency of the first signal, $\omega_1$. Monotonic dependence is discernible. (e) Neural states $\bf x$ in the PC space at the end of the delay period ($t=T_f$) for $\omega_1=1.5, 3, 4.5$ with various $0\leq \phi \leq \pi$. (f) Directions of neural trajectories driven by the second signal with $\omega_2=3$. The x-axis shows the amplitude of the trajectory at $t=T_f$. Arrows show the direction and magnitude of the change in the PC3 component of ${\bf x}(T_f+1)-{\bf x}(T_f)$ for the first input signal with $\omega_1=1.5,2,2.5,3.5,4,4.5$. 
        }
\end{figure*}

Here, the neural activities continued to change over time during the delay period, as illustrated in \footnote{See Supplemental Material Fig.1 for the characteristics of the neural activities during the delay period.}. It can now be confirmed that the first signal was memorized by transient dynamics rather than by the attractor. 
To further confirm this, the second signal was removed, and the long-term behavior of neural dynamics after the delay period was observed. 
As shown in Fig.\ref{fig:fig3}a, when the delay period was prolonged, the neural states converged to a limit-cycle attractor independent of the signal frequency, $\omega_1$. 
After the first signal input, neural activities were attracted to a low-dimensional manifold at which the limit cycle was located\footnote{The fixed-point analysis\cite{barak_fixed_point} showed that there were a few slow points acting as pseudo-saddles through which the neural activity passed after the first signal. See Supplemental Material Fig.3 for the slow points plotted in the 3-dimensional PC space.}. Then, the neural activities oscillated with slowly increasing amplitude towards the limit-cycle attractor. This increase, however, was so slow that the short-term memory was maintained for a sufficient amount of time. We analyze the duration of memory in detail below.

Subsequently, we show how memory information (i.e., frequency of signals) was represented in the neural activity. 
In Fig.\ref{fig:fig3}a, we can see the trend in which the amplitude of the transient oscillation before the attraction to the limit cycle has monotonic dependence on $\omega_1$. At the end of the delay period ($t=T_f$), this trend is remarkable; there is strong negative correlation between the amplitude, $|{\bf x}(t=T_f)|$, and the first signal frequency, $\omega_1$(Fig.\ref{fig:fig3}d2). Hence, $\omega_1$ is encoded by the amplitude of the transient oscillation(Fig.\ref{fig:fig3}c). 
Notably, this monotonic coding of the input frequency by the amplitude does not hold immediately after the input of the first signal. Indeed, at the beginning of the delay period ($t=T_s$), there was no such monotonic dependence on $\omega_1$ (see Fig.\ref{fig:fig3}b, d1). Therefore the neural dynamics during the delay period ($T_s\leq t\leq T_f)$ shaped the manifold from Fig.\ref{fig:fig3}b to Fig.\ref{fig:fig3}c. 

This coding by amplitude is beneficial for discarding information irrelevant to the task. In the present task, the input signals included phases of the oscillation apart from the frequency. Hence the RNN should discriminate the frequency information from the phase information. Here, signals having different phases were mapped to different points on the same radius trajectory (Fig.\ref{fig:fig3}e). The neural dynamics during the delay period ($T_s\leq t\leq T_f)$ dampened information on the signal other than its frequency, $\omega_1$. 

The neural state corresponding to $\omega_1$, encoded by the amplitude of transient oscillation, was used as the initial state for the response to the second signal. After the second signal was input, the neural states moved separately up and down along the PC3-axis according to the sizes of $\omega_2$ and $\omega_1$. If $\omega_2 > \omega_1$, it moved in the positive direction along the PC3-axis, and vice versa. In Fig.\ref{fig:fig3}f, changes in the PC3 component of ${\bf x}(t)$ after the second signal with $\omega_2=3$ are plotted against the amplitude of ${\bf x}(t)$ oscillation at $t=T_f$.
As shown, depending on the amplitude of the oscillation at $t=T_f$, PC3 moved upwards when $|{\bf x}(t=T_f)|$ was small (corresponding to $\omega_2<\omega_1$), and it moved downwards when $|{\bf x}(t=T_f)|$ was large (corresponding to $\omega_2>\omega_1$).
This property separated the neural states in the direction of PC3, as shown in Fig.\ref{fig:fig2}b, and it enabled the RNN to correctly solve the frequency comparison task by using the short-term memory for $\omega_1$, as coded in the transient amplitude.

To verify the generality of short-term memory encoded by the amplitude of transient oscillation, we examined two other settings. First, we changed the length of the delay period as follows. During the training phase, $T_d$ was homogeneously distributed as $T_d \in [75,105]$, and during the test phase, $T_d$ was fixed at 90. With this longer delay period setting, we confirmed that the present mechanism also worked \footnote{See Supplemental Material Fig.2 for the neural activity of trained RNN with longer delay period.}. Second, we trained the RNN for two other comparison tasks, which requested we compare the velocity and noise variance. In the former task, the first and second signals were given by $u_{1,2}(t)=a_{1,2}t+b+\eta_{1,2}(t)$, and the RNN was required to determine which of the velocities, $a_{1,2}$, was larger. 
In the latter task, the first and second signals were given by $u_{1,2}=\sin(t+\phi )+\eta_{1,2}(t)$, where $\eta_{1,2}(t)$ was a random Gaussian variable having an average of zero and a standard deviation of $\sigma_{\rm ex}^{1,2}$. 
The RNN was then required to determine which of the variances, $\sigma_{\rm ex}^{1,2}$, was larger. 
For both the tasks, a monotonic dependence between the L2 norm of the neural states (the amplitude of transient oscillation) and the parameter to be compared was revealed at the end of the delay period ($t=T_f$)\footnote{See Supplemental Material Fig.4 and 5 for the results of the different tasks.}. The information of the velocity, $a_1$, (for the former) and the noise variance, $\sigma_{\rm ex}^1$, (for the latter) was memorized as the amplitude of transient oscillation.

\subsection{Convergence to the limit cycle}

As described previously, transient oscillatory dynamics eventually converged to the limit cycle. 
When this trajectory converged, the memory was forgotten, and the information of the input was lost. 
Hence, for this memory to work and to accomplish the task, the time to converge to the limit cycle must be sufficiently long. As a measure of convergence of ${\bf x}(t)$ to the limit cycle, we first introduced $\ell(t)=|{\bf x}(t)-{\bf x}(t+T_L)|$, where $T_L$ is the period of the limit cycle. Subsequently, we estimated the convergence time to the limit cycle as the time, $t$, when $\ell(t)$ was sufficiently small (i.e., the time $t$ at which $\ell(t)$ first satisfies $\ell(t) \leq 0.05 $ for the first time). We computed the average convergence time over 100 trajectories, ${\bf x}(t)$, for different $\omega_1$ values and noted that the convergence time was a dozen times longer than $T_f$. We also found that during training, the convergence time increased alongside an increase in the accuracy of the task \footnote{See Supplemental Material Fig.6 for the convergence time to the limit cycle.}.
These results suggest that the short-term memory was maintained over a sufficiently long period, prolonged after the delay period, and the length of the convergence time was related to the performance of informational processing.

\subsection{Robustness to noise}
As described in the introduction, the mechanism of the robustness to noise in the short-term memory by transient oscillation remains elusive. 
Hence, the robustness of the memory was examined by adding the noise term in Eq.(1) using the Langevin equation:

\begin{equation}
    \dot{x}_i = -x_i + \sum_{j=1}^N J_{ij}\tanh(x_j)+W_i^{\rm in}u + \xi_i,
\end{equation}
where $\xi_i$ is a random Gaussian variable with an average of zero and a standard deviation of $\sigma_{\rm neu}$. As with the no-noise setting, we discretized the dynamics and obtained the following equation: $x_i(t+1)  = (1-\alpha)x_i(t)+ \alpha (\sum_{j=1}^N J_{ij}\tanh(x_j(t))+W_i^{in}u(t)) +\sqrt{\alpha}\xi_i(t)$. The RNN was trained under a noise level, $\sigma_{\rm neu}^{\rm train}$, with regularization for the average squared norm of neural activity, $\frac{1}{T_a}\sum_t^{T_a} \sum_i^N x_i(t)^2$. 
Regularization was applied to the loss function in the form of 

\begin{equation}
    \mathcal L_{\rm noise} = {\mathcal L} + \lambda_{\rm act}\frac{1}{T_a}\sum_t^{T_a} \sum_i^N x_i(t)^2.
\end{equation}

The reason that we introduced this regularization term is as follows. Notably, if the average squared norm, $\frac{1}{T_a}\sum_t^{T_a} \sum_i^N x_i(t)^2$, increases, robustness will increase because the derivative of $\tanh(x)$ will generally decrease. 
By using this regularization term to suppress the norm of neural activity, we can avoid such trivial robustness. We experimentally determined $\lambda_{\rm act}$ so that the norm of the internal dynamics of RNNs trained with noise ($\sigma_{\rm neu}^{\rm train}=0.04$) would be comparable to those trained without noise ($\sigma_{\rm neu}^{\rm train}=0$), and we adopted $\lambda_{\rm act}=30$. Subsequently, the trained network was tested for solving the task under the noise level, $\sigma_{\rm neu}^{\rm test}$.

\begin{figure}
    \includegraphics[width=7cm]{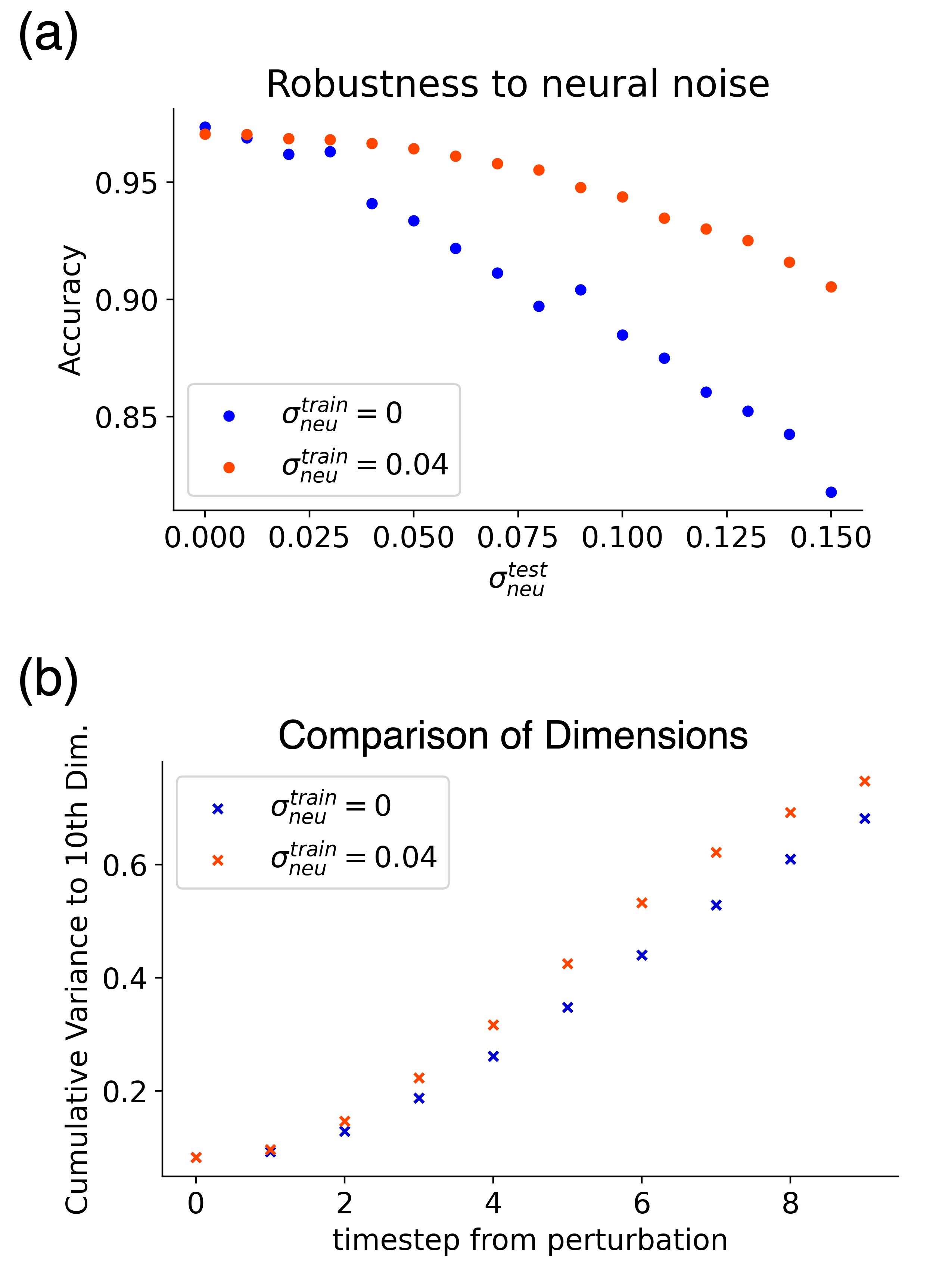}
    \caption{
        \label{fig:fig4} (a) Accuracy of the frequency comparison task plotted against the noise level, $\sigma_{\rm neu}^{\rm test}$. Different colors show the RNN trained at different levels of noise ($\sigma_{\rm neu}^{\rm train}=0, 0.04)$. Bold lines show the mean score, and shaded areas show the standard deviations. (b) The cumulative contribution ratio of the perturbed trajectories up to the 10th principal component (i.e. the cumulative percentage of the eigenvalues corresponding to 1st-to-10th eigenvectors $(\sum_{i=1}^{10} \lambda_i)/(\sum_{i=1}^{256} \lambda_i)$. These are plotted against the time following the application of perturbation for $\sigma_{\rm neu}^{\rm train}=0$(blue), and $\sigma_{\rm neu}^{\rm train}=0.04$(red). If it is larger, more points are restricted within the lower-dimensional manifold. Because the perturbation is completely random, the perturbation dimension is high immediately after it is applied. However, over time, it is attracted to a lower-dimensional manifold, especially for $\sigma_{\rm neu}^{\rm train}=0.04$.
        }
\end{figure}

As depicted in Fig.\ref{fig:fig4}a, the accuracy of the short-term memory task decreased with the applied noise, $\sigma_{\rm neu}^{\rm test}$, for the network trained without noise. In contrast, for the model trained with a sufficient noise level, $\sigma_{\rm neu}^{\rm train}$, the drop in the accuracy by the increase in noise level was suppressed, even up to a noise level higher than $\sigma_{\rm neu}^{\rm train}$. The system gained a higher robustness to noise, which was beyond the level added during learning. Notably, the RNN trained with neural noise realized the same mechanism of the short-term memory with transient oscillation.

To examine the achievement of the stability of short-term memory, the following perturbation analysis was performed. We set $\sigma_{\rm neu}^{\rm test}=0$, and subsequently, we introduced an instantaneous perturbation during the delay period (i.e., $T_s \leq t_{\rm per}\leq T_f$, as ${\bf x}_{\rm per}(t) = {\bf x}(t)+\bf \delta x\delta_{t,t_{\rm per}}$), where ${\bf \delta x}$ is a random variable that follows a Gaussian distribution of mean zero and of variance $\sigma_{\rm per}^2$, and $\delta_{t,t_{\rm per}}$ represents a Kronecker delta.

Intuitively, one might expect that the distance between the perturbed trajectory and the original trajectory would be decreased for the networks trained under noise. Here, we first computed the L2 norm distance between the perturbed and original trajectories: $L_{per} = |{\bf x}_{per}(t) - {\bf x}(t)|$. 
In contrast to the naive expectation, however, the distance does not depend substantially on the trained noise level, $\sigma_{\rm neu}^{\rm train}$. 
Hence, the distance between the trajectories cannot explain the memory robustness for a system trained with noise $\sigma_{\rm neu}^{\rm train}$. 
To understand the robustness, the direction of the perturbation of the trajectories must be considered. 
Indeed, the shift of the trajectory along the original trajectory was not harmful to the memory discussed, because the amplitude of the transient oscillation was not affected by the shift. 
Hence, the dimension of the manifold spanned by the perturbed trajectories is more important to the robustness of memory than the distances between the perturbed trajectories and the original trajectories.

\begin{figure}
    \includegraphics[width=7cm]{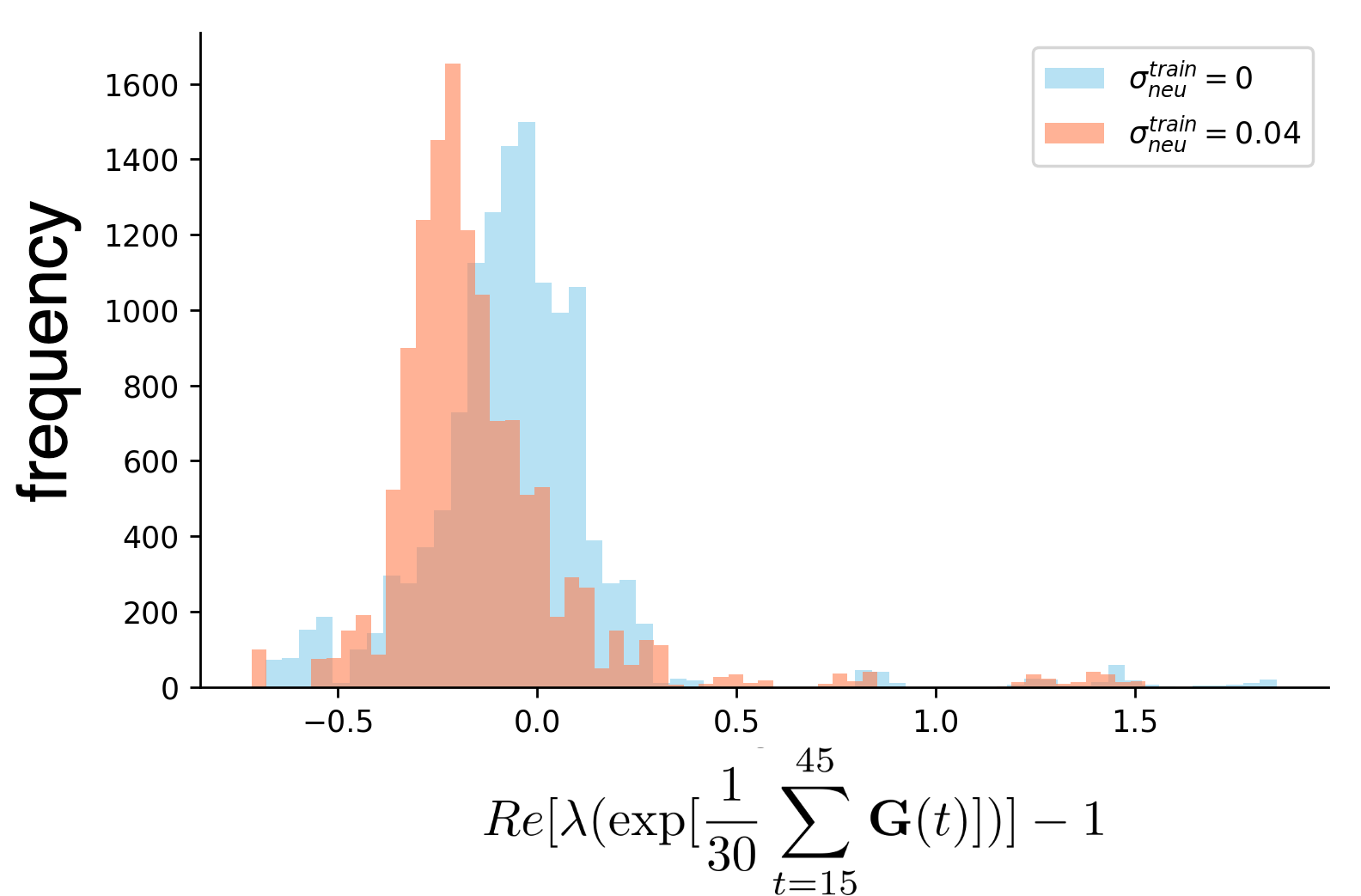}
    \caption{
        \label{fig:jacobian_eigenvalues}Histogram of the frequency distribution of the real part of the eigenvalues of $\exp[\frac{1}{30}\sum_{t=15}^{45} {\bf G}(t)]$. We calculated ${\bf G}(t)$ for 50 samples of the input signals. Notably, there were 256 eigenvalues for one $\exp[\frac{1}{30}\sum_{t=15}^{45} {\bf G}(t)]$. Blue is the result for the RNN trained without noise, and red is the result for the RNN trained with noise($\sigma_{\rm neu}^{\rm train}=0.04$). In the latter, the eigenvalues were biased to the negative side overall.
    }
\end{figure}

Accordingly, we estimated the dimension of the manifold spanned by a large number of perturbed trajectories. We adopted a principal component analysis (PCA) for an ensemble of these perturbed trajectories. If perturbation causes a shift mainly along the trajectory direction, perturbed trajectories should be restricted within a low-dimensional manifold along the original trajectory. Hence, the dimension of perturbed trajectories estimated by PCA will work as a measure of robustness in the directions of perturbed trajectories. To estimate the dimension, we computed the percentage of the variance corresponding the 1st-to-10th eigenvectors for each time after the perturbation as $(\sum_{i=1}^{10} \lambda_i)/(\sum_{i=1}^{256} \lambda_i)$, where $\lambda_i$ represents the eigenvalue of covariance matrix with an ordering to satisfy $\lambda_1 \geq \lambda_2 \geq ... \geq \lambda_{256}$.

This percentage was larger when the perturbed trajectory was restricted to a lower-dimensional manifold. As shown in Fig. 4, for the RNN trained with noise $\sigma_{neu}$, the trajectory falls into a lower-dimensional manifold with over time following the perturbation. The percentage for such robust RNN is larger than the RNN trained without neural noise. This result implies that the dynamics after the perturbation were more restricted to a lower-dimensional manifold for the model trained with noise. 
This low-dimensional compression is key to the robustness of the neural noise. 
Notably, another method of estimating the attractor dimension was proposed in \cite{SIROVICH1989126, Ciliberto_1991}, where the number of eigenvectors of the principal components required to achieve 90\% of the total variance was calculated. We have confirmed that similar results have been obtained by applying this method \footnote{See Supplemental Material Fig.7 for the estimated dimension of perturbed trajectories by the method in \cite{SIROVICH1989126, Ciliberto_1991}}.

To characterize the mechanism of this convergence of the transient trajectory to low-dimensional space, we focused on the Jacobi matrix, ${\bf G}(t)$, of this system. The Jacobi matrix is defined as the following equation:

\begin{equation}
    G_{ij}(t) = J_{ij}\tanh'(x_j(t)).
\end{equation}

The perturbed dynamics convergence can be estimated by the eigenvalues of matrix $e^{\frac{1}{T}\int dt {\bf G}(t)}$ along the trajectory, as was adopted in the calculation of finite-time Lyapunov exponents\cite{lyapunov}. Therefore, we investigated the difference of eigenvalues between the RNNs trained with $\sigma_{\rm neu}^{\rm train}=0$ and those trained with $\sigma_{\rm neu}^{\rm train}=0.04$. Because the value of the Jacobi matrix depends on the specific trajectory, ${\bf x}(t)$, we calculated the histograms of the eigenvalues for 50 samples of input signals (Fig.\ref{fig:jacobian_eigenvalues}). We also adopted the discretization to calculate $e^{\frac{1}{T}\int dt {\bf G}(t)}$ as $\exp[\frac{1}{30}\sum_{t=15}^{45} {\bf G}(t)]$.

For the latter RNN, there were more eigenvalues with negative real parts. As shown in Fig.\ref{fig:jacobian_eigenvalues}, the number of eigenvalues with positive real parts was reduced. This result suggests that the restriction to a low-dimensional manifold in the robust RNN was caused by the compression of many modes of perturbed dynamics. This robustness caused by the contraction of transient trajectories was clearly distinguishable from that caused by the stability of fixed-point attractors.

\subsection{Different types of neural networks}
To investigate the range of validity of transient oscillatory short-term memory, we considered two different types of neural networks. First, we considered neural networks in which synapses obey Dale's law (i.e., the neural network consists of excitatory and inhibitory neurons, and synapses extended from a given neuron are either all excitatory , or inhibitory)\cite{Dale, 10.1371/journal.pcbi.1004792}. Second, we considered neural networks in which synapses were sparse. We call these networks excitatory-inhibitory and Sparse networks, respectively.

We trained the RNN to solve the frequency comparison task while satisfying each condition. Under the excitatory-inhibitory networks condition, as in \cite{Efficient}, hidden neurons were divided into excitatory and inhibitory neurons in a 4-to-1 ratio. $\bf J$ was constrained so it can be decomposed as ${\bf J}={\bf J}^+ {\bf D}$, where ${\bf J}^+$ is a matrix with all components greater than or equal to zero, and $\bf D$ is a diagonal matrix that satisfies $D_{ii}=1$ if $i\leq 180$ and $D_{ii}=-1$ if $i> 180$. With this constraint, it can be said that $\bf J$ obeys Dale's law. During the training phase, ${\bf D}$ was fixed, and ${\bf W}^{\rm in}$, ${\bf J}^+$ and ${\bf W}^{\rm out}$ were adjusted. In Sparse networks, we set the percentage of synapses with nonzero weight to approximately 20\%. We adopted the deep-rewiring technique\cite{deep-r} to train under constraints where only 20\% of synapses have nonzero values, and all others are zero. As in Supplementary Material Figs. 8 and 9, after learning, the RNNs solved the frequency comparison task by maintaining short-term memory by transient oscillatory dynamics \footnote{See Supplemental Material Fig.8 and 9 for the results under the excitatory-inhibitory networks condition and the sparse networks condition}.

\section{Discussion}
In this study, we uncovered a generic scheme for short-term memory sustained by transient oscillatory dynamics by training RNNs to achieve a task requiring short-term memory to compare two interspaced input signals\cite{versatile,Richards2019}. 
We demonstrated that short-term memory was encoded in the amplitude of transient oscillations of neural activities. 
The neural state given by high-dimensional dynamical systems fell into a low-dimensional manifold\cite{barak_fixed_point} and exhibited transient oscillation, which slowly approached a limit-cycle attractor. 
With the passage of time, continuous information from the first signal input, such as the frequency and velocity of the input signal, was encoded in the amplitude of the transient oscillation and maintained as short-term memory.
Other irrelevant information in the inputs (e.g., phase and noise) was discarded during the transient dynamics. 
Hence, short-term memory is encoded and maintained by transient dynamics and is robust to external noise.

The proposed mechanism, wherein the memory is encoded in transient oscillation, contrasts with the view of memory as fixed-point attractors (i.e., memories as attractors). 
When the RNN is trained to store discrete information (e.g., the possible input signal frequency candidates are limited to 1, 2, 3, 4, or 5 Hz), the memories are encoded by multiple fixed-point attractors (i.e., persistent activities)\cite{romo_model}. 
In the task adopted in this study, by contrast, storing continuous information was needed. Hence, encoding into the amplitude of transient oscillation occurred. Notably, as an alternative mechanism to encoding continuous information, a line attractor was proposed\cite{Seung13339, Mante2013}. 
Here, attractors coexisted continuously on a line in the state space of neural activities, along which the fixed-point was marginally stable (i.e., one of the eigenvalues of the Jacobi matrix (with the eigenvector along the line) should be zero). 
However, in autonomous dynamical systems, such marginally stable attractors are not generic, and for their existence, special constraints are required. 
Indeed, in this study, such line attractors were not shaped by learning\footnote{For the possible applicability of chaotic attractors to short-term memory, see J. S. Nicolis and I. Tsuda, Chaotic dynamics of information processing:  The ``magic  number  seven  plus-minus two’’ revisited, Bulletin of Mathematical Biology47, 343(1985).}.

As another mechanism of short-term memory, synaptic plasticity has been proposed\cite{Mongillo1543, Tsodyks2017, 10.1371}. In this case, the synaptic connections are modeled as time-varying not only during training, but also during inferencing. The memory information is encoded in the time-varying strength of the synapses. 
In this study, because we focused on neural-based short-term memory, synaptic connections (i.e., the vector ${\bf W}^{\rm in}$ and two matrices ${\bf J}$ and ${\bf W}^{\rm out}$) change only during training and were fixed thereafter. 
In future research, we plan to investigate how short-term memory based on transient oscillation can be realized when considering synaptic plasticity after training.

It remains to be seen if the present scheme for short-term memory can be adopted biologically. 
The results suggest that the short-term memory encoded by the amplitude of transient oscillation works over a wide range of systems. This may be plausible, because the present mechanism is generally represented in terms of dynamical systems, and it is not difficult to generate oscillatory dynamics from an ensemble of neurons. 
Here, we offer two remarks. We expect that the present scheme is also valid therein, as it is robust to noise in neural dynamics. 
Next, oscillatory dynamics from certain modes of neural activities evoked by inputs were observed from neural data references\cite{travellingwave, travellingwave2}. Notably, such dynamics are not necessarily observed just by computing the average neural activity, which is often a rather stationary reference independent of inputs. Indeed, this is true in the proposed model. The prominent oscillation depending on the input, as shown in Fig.2, is observed only by taking appropriate principal components. If we compute just the average activity over all neurons, it is almost stationary, and the oscillation is difficult to discern, which is consistent with experimental observations.

In this study, attraction to the low-dimensional manifold of perturbation allowed for the transient trajectory to be robust to noise. 
Recently, dynamic robustness has been regarded as an important property of information processing in the brain\cite{transient_robustness}. The present results provide general insight into the mechanism of this type of robustness. Furthermore, in cell and developmental systems, such dynamic robustness has also been discussed as homeorhesis\cite{Waddington,Matsushita,Jon}, and the results may be applicable to such phenomena. 

\begin{acknowledgments}
The authors would like to express their gratitude to Tomoki Kurikawa, Ichiro Tsuda, Hiromichi Suetani, and Tetsuhiro S. Hatakeyama for their comments and discussions. This research was partially supported by a Grant-in-Aid for Scientific Research (A) (20H00123) from the Japanese Society for the Promotion of Science (JSPS).
\end{acknowledgments}

\bibliography{reference}

\providecommand{\noopsort}[1]{}\providecommand{\singleletter}[1]{#1}%
\begin{thebibliography}{58}%
\makeatletter
\providecommand \@ifxundefined [1]{%
 \@ifx{#1\undefined}
}%
\providecommand \@ifnum [1]{%
 \ifnum #1\expandafter \@firstoftwo
 \else \expandafter \@secondoftwo
 \fi
}%
\providecommand \@ifx [1]{%
 \ifx #1\expandafter \@firstoftwo
 \else \expandafter \@secondoftwo
 \fi
}%
\providecommand \natexlab [1]{#1}%
\providecommand \enquote  [1]{``#1''}%
\providecommand \bibnamefont  [1]{#1}%
\providecommand \bibfnamefont [1]{#1}%
\providecommand \citenamefont [1]{#1}%
\providecommand \href@noop [0]{\@secondoftwo}%
\providecommand \href [0]{\begingroup \@sanitize@url \@href}%
\providecommand \@href[1]{\@@startlink{#1}\@@href}%
\providecommand \@@href[1]{\endgroup#1\@@endlink}%
\providecommand \@sanitize@url [0]{\catcode `\\12\catcode `\$12\catcode
  `\&12\catcode `\#12\catcode `\^12\catcode `\_12\catcode `\%12\relax}%
\providecommand \@@startlink[1]{}%
\providecommand \@@endlink[0]{}%
\providecommand \url  [0]{\begingroup\@sanitize@url \@url }%
\providecommand \@url [1]{\endgroup\@href {#1}{\urlprefix }}%
\providecommand \urlprefix  [0]{URL }%
\providecommand \Eprint [0]{\href }%
\providecommand \doibase [0]{http://dx.doi.org/}%
\providecommand \selectlanguage [0]{\@gobble}%
\providecommand \bibinfo  [0]{\@secondoftwo}%
\providecommand \bibfield  [0]{\@secondoftwo}%
\providecommand \translation [1]{[#1]}%
\providecommand \BibitemOpen [0]{}%
\providecommand \bibitemStop [0]{}%
\providecommand \bibitemNoStop [0]{.\EOS\space}%
\providecommand \EOS [0]{\spacefactor3000\relax}%
\providecommand \BibitemShut  [1]{\csname bibitem#1\endcsname}%
\let\auto@bib@innerbib\@empty
\bibitem [{\citenamefont {Jonides}\ \emph {et~al.}(2008)\citenamefont
  {Jonides}, \citenamefont {Lewis},\ and\ \citenamefont
  {et~al.}}]{review-of-short-term-memory}%
  \BibitemOpen
  \bibfield  {author} {\bibinfo {author} {\bibfnamefont {J.}~\bibnamefont
  {Jonides}}, \bibinfo {author} {\bibfnamefont {R.~L.}\ \bibnamefont {Lewis}},
  \ and\ \bibinfo {author} {\bibnamefont {et~al.}},\ }\href@noop {} {\bibfield
  {journal} {\bibinfo  {journal} {Annual Review of Psychology}\ }\textbf
  {\bibinfo {volume} {59}},\ \bibinfo {pages} {193} (\bibinfo {year}
  {2008})}\BibitemShut {NoStop}%
\bibitem [{\citenamefont {Colby}\ \emph {et~al.}(1996)\citenamefont {Colby},
  \citenamefont {Duhamel},\ and\ \citenamefont
  {Goldberg}}]{doi:10.1152/jn.1996.76.5.2841}%
  \BibitemOpen
  \bibfield  {author} {\bibinfo {author} {\bibfnamefont {C.~L.}\ \bibnamefont
  {Colby}}, \bibinfo {author} {\bibfnamefont {J.~R.}\ \bibnamefont {Duhamel}},
  \ and\ \bibinfo {author} {\bibfnamefont {M.~E.}\ \bibnamefont {Goldberg}},\
  }\href {\doibase 10.1152/jn.1996.76.5.2841} {\bibfield  {journal} {\bibinfo
  {journal} {Journal of Neurophysiology}\ }\textbf {\bibinfo {volume} {76}},\
  \bibinfo {pages} {2841} (\bibinfo {year} {1996})},\ \bibinfo {note} {pMID:
  8930237},\ \Eprint
  {http://arxiv.org/abs/https://doi.org/10.1152/jn.1996.76.5.2841}
  {https://doi.org/10.1152/jn.1996.76.5.2841} \BibitemShut {NoStop}%
\bibitem [{\citenamefont {Funahashi}\ \emph {et~al.}(1989)\citenamefont
  {Funahashi}, \citenamefont {Bruce},\ and\ \citenamefont
  {Goldman-Rakic}}]{Funahashi1989}%
  \BibitemOpen
  \bibfield  {author} {\bibinfo {author} {\bibfnamefont {S.}~\bibnamefont
  {Funahashi}}, \bibinfo {author} {\bibfnamefont {C.}~\bibnamefont {Bruce}}, \
  and\ \bibinfo {author} {\bibfnamefont {P.}~\bibnamefont {Goldman-Rakic}},\
  }\href@noop {} {\bibfield  {journal} {\bibinfo  {journal} {Journal of
  neurophysiology}\ }\textbf {\bibinfo {volume} {61}},\ \bibinfo {pages} {331}
  (\bibinfo {year} {1989})}\BibitemShut {NoStop}%
\bibitem [{\citenamefont {Fuster}\ and\ \citenamefont
  {Alexander}(1971)}]{Fuster1971}%
  \BibitemOpen
  \bibfield  {author} {\bibinfo {author} {\bibfnamefont {J.}~\bibnamefont
  {Fuster}}\ and\ \bibinfo {author} {\bibfnamefont {G.}~\bibnamefont
  {Alexander}},\ }\href@noop {} {\bibfield  {journal} {\bibinfo  {journal}
  {Science}\ }\textbf {\bibinfo {volume} {173}},\ \bibinfo {pages} {652}
  (\bibinfo {year} {1971})}\BibitemShut {NoStop}%
\bibitem [{\citenamefont {Wang}(2001)}]{fixed-point-short-term-memory}%
  \BibitemOpen
  \bibfield  {author} {\bibinfo {author} {\bibfnamefont {X.~J.}\ \bibnamefont
  {Wang}},\ }\href {\doibase 10.1016/S0166-2236(00)01868-3} {\bibfield
  {journal} {\bibinfo  {journal} {Trends in Neurosciences}\ }\textbf {\bibinfo
  {volume} {24}},\ \bibinfo {pages} {455} (\bibinfo {year} {2001})}\BibitemShut
  {NoStop}%
\bibitem [{\citenamefont {Haider}\ and\ \citenamefont
  {McCormick}(2009)}]{HAIDER2009171}%
  \BibitemOpen
  \bibfield  {author} {\bibinfo {author} {\bibfnamefont {B.}~\bibnamefont
  {Haider}}\ and\ \bibinfo {author} {\bibfnamefont {D.~A.}\ \bibnamefont
  {McCormick}},\ }\href {\doibase https://doi.org/10.1016/j.neuron.2009.04.008}
  {\bibfield  {journal} {\bibinfo  {journal} {Neuron}\ }\textbf {\bibinfo
  {volume} {62}},\ \bibinfo {pages} {171} (\bibinfo {year} {2009})}\BibitemShut
  {NoStop}%
\bibitem [{\citenamefont {Baeg}\ \emph {et~al.}(2003)\citenamefont {Baeg},
  \citenamefont {Kim}, \citenamefont {Huh}, \citenamefont {Mook-Jung},
  \citenamefont {Kim},\ and\ \citenamefont
  {Jung}}]{Dynamics_of_Population_Code}%
  \BibitemOpen
  \bibfield  {author} {\bibinfo {author} {\bibfnamefont {E.~H.}\ \bibnamefont
  {Baeg}}, \bibinfo {author} {\bibfnamefont {Y.~B.}\ \bibnamefont {Kim}},
  \bibinfo {author} {\bibfnamefont {K.}~\bibnamefont {Huh}}, \bibinfo {author}
  {\bibfnamefont {I.}~\bibnamefont {Mook-Jung}}, \bibinfo {author}
  {\bibfnamefont {H.~T.}\ \bibnamefont {Kim}}, \ and\ \bibinfo {author}
  {\bibfnamefont {M.~W.}\ \bibnamefont {Jung}},\ }\bibfield  {booktitle} {\emph
  {\bibinfo {booktitle} {Neuron}},\ }\href {\doibase
  10.1016/S0896-6273(03)00597-X} {\bibfield  {journal} {\bibinfo  {journal}
  {Neuron}\ }\textbf {\bibinfo {volume} {40}},\ \bibinfo {pages} {177}
  (\bibinfo {year} {2003})}\BibitemShut {NoStop}%
\bibitem [{\citenamefont {Bondanelli}\ and\ \citenamefont
  {Ostojic}(2020)}]{ostojic2020}%
  \BibitemOpen
  \bibfield  {author} {\bibinfo {author} {\bibfnamefont {G.}~\bibnamefont
  {Bondanelli}}\ and\ \bibinfo {author} {\bibfnamefont {S.}~\bibnamefont
  {Ostojic}},\ }\href@noop {} {\bibfield  {journal} {\bibinfo  {journal} {PLoS
  Comput Biol}\ }\textbf {\bibinfo {volume} {16}} (\bibinfo {year}
  {2020})}\BibitemShut {NoStop}%
\bibitem [{\citenamefont {Druckmann}\ and\ \citenamefont
  {Chklovskii}(2012)}]{Druckmann2012}%
  \BibitemOpen
  \bibfield  {author} {\bibinfo {author} {\bibfnamefont {S.}~\bibnamefont
  {Druckmann}}\ and\ \bibinfo {author} {\bibfnamefont {D.}~\bibnamefont
  {Chklovskii}},\ }\href {\doibase 10.1016/j.cub.2012.08.058} {\bibfield
  {journal} {\bibinfo  {journal} {Current biology : CB}\ }\textbf {\bibinfo
  {volume} {22}} (\bibinfo {year} {2012}),\
  10.1016/j.cub.2012.08.058}\BibitemShut {NoStop}%
\bibitem [{\citenamefont {Goldman}(2009)}]{Goldman2009}%
  \BibitemOpen
  \bibfield  {author} {\bibinfo {author} {\bibfnamefont {M.~S.}\ \bibnamefont
  {Goldman}},\ }\href {\doibase 10.1016/j.neuron.2008.12.012} {\bibfield
  {journal} {\bibinfo  {journal} {Neuron}\ }\textbf {\bibinfo {volume} {61}},\
  \bibinfo {pages} {621} (\bibinfo {year} {2009})}\BibitemShut {NoStop}%
\bibitem [{\citenamefont {Orhan}\ and\ \citenamefont
  {Pitkow}(2020)}]{Orhan2020Improved}%
  \BibitemOpen
  \bibfield  {author} {\bibinfo {author} {\bibfnamefont {E.}~\bibnamefont
  {Orhan}}\ and\ \bibinfo {author} {\bibfnamefont {X.}~\bibnamefont {Pitkow}},\
  }in\ \href {https://openreview.net/forum?id=ryx1wRNFvB} {\emph {\bibinfo
  {booktitle} {International Conference on Learning Representations}}}\
  (\bibinfo {year} {2020})\BibitemShut {NoStop}%
\bibitem [{\citenamefont {Orhan}\ and\ \citenamefont {Ma}(2019)}]{Orhan2019}%
  \BibitemOpen
  \bibfield  {author} {\bibinfo {author} {\bibfnamefont {E.}~\bibnamefont
  {Orhan}}\ and\ \bibinfo {author} {\bibfnamefont {W.~J.}\ \bibnamefont {Ma}},\
  }\href@noop {} {\bibfield  {journal} {\bibinfo  {journal} {Nature
  Neuroscience}\ }\textbf {\bibinfo {volume} {22}},\ \bibinfo {pages} {275}
  (\bibinfo {year} {2019})}\BibitemShut {NoStop}%
\bibitem [{\citenamefont {Rabinovich}\ \emph {et~al.}(2008)\citenamefont
  {Rabinovich}, \citenamefont {Huerta},\ and\ \citenamefont
  {Laurent}}]{Rabinovich48}%
  \BibitemOpen
  \bibfield  {author} {\bibinfo {author} {\bibfnamefont {M.}~\bibnamefont
  {Rabinovich}}, \bibinfo {author} {\bibfnamefont {R.}~\bibnamefont {Huerta}},
  \ and\ \bibinfo {author} {\bibfnamefont {G.}~\bibnamefont {Laurent}},\ }\href
  {\doibase 10.1126/science.1155564} {\bibfield  {journal} {\bibinfo  {journal}
  {Science}\ }\textbf {\bibinfo {volume} {321}},\ \bibinfo {pages} {48}
  (\bibinfo {year} {2008})}\BibitemShut {NoStop}%
\bibitem [{\citenamefont {Sussillo}\ and\ \citenamefont
  {Barak}(2013)}]{barak_fixed_point}%
  \BibitemOpen
  \bibfield  {author} {\bibinfo {author} {\bibfnamefont {D.}~\bibnamefont
  {Sussillo}}\ and\ \bibinfo {author} {\bibfnamefont {O.}~\bibnamefont
  {Barak}},\ }\href@noop {} {\bibfield  {journal} {\bibinfo  {journal} {Naural
  Comput.}\ }\textbf {\bibinfo {volume} {25(3)}},\ \bibinfo {pages} {626}
  (\bibinfo {year} {2013})}\BibitemShut {NoStop}%
\bibitem [{\citenamefont {Maheswaranathan}\ \emph {et~al.}(2019)\citenamefont
  {Maheswaranathan}, \citenamefont {Williams},\ and\ \citenamefont
  {et~al}}]{Sussillo2019}%
  \BibitemOpen
  \bibfield  {author} {\bibinfo {author} {\bibfnamefont {N.}~\bibnamefont
  {Maheswaranathan}}, \bibinfo {author} {\bibfnamefont {A.}~\bibnamefont
  {Williams}}, \ and\ \bibinfo {author} {\bibnamefont {et~al}},\ }in\
  \href@noop {} {\emph {\bibinfo {booktitle} {Advances in Neural Information
  Processing Systems 32}}}\ (\bibinfo  {publisher} {Neural Information
  Processing Systems Foundation},\ \bibinfo {year} {2019})\ pp.\ \bibinfo
  {pages} {15629--15641}\BibitemShut {NoStop}%
\bibitem [{\citenamefont {Romo}\ \emph {et~al.}(1999)\citenamefont {Romo},
  \citenamefont {Brody}, \citenamefont {Hernández},\ and\ \citenamefont
  {L.}}]{romo}%
  \BibitemOpen
  \bibfield  {author} {\bibinfo {author} {\bibfnamefont {R.}~\bibnamefont
  {Romo}}, \bibinfo {author} {\bibfnamefont {C.}~\bibnamefont {Brody}},
  \bibinfo {author} {\bibfnamefont {A.}~\bibnamefont {Hernández}}, \ and\
  \bibinfo {author} {\bibfnamefont {L.}~\bibnamefont {L.}},\ }\href@noop {}
  {\bibfield  {journal} {\bibinfo  {journal} {Nature}\ }\textbf {\bibinfo
  {volume} {399}},\ \bibinfo {pages} {470} (\bibinfo {year}
  {1999})}\BibitemShut {NoStop}%
\bibitem [{\citenamefont {Rumelhart}\ \emph {et~al.}(1986)\citenamefont
  {Rumelhart}, \citenamefont {Hinton},\ and\ \citenamefont
  {Williams}}]{Rumelhart:1986we}%
  \BibitemOpen
  \bibfield  {author} {\bibinfo {author} {\bibfnamefont {D.~E.}\ \bibnamefont
  {Rumelhart}}, \bibinfo {author} {\bibfnamefont {G.~E.}\ \bibnamefont
  {Hinton}}, \ and\ \bibinfo {author} {\bibfnamefont {R.~J.}\ \bibnamefont
  {Williams}},\ }\href {\doibase 10.1038/323533a0} {\bibfield  {journal}
  {\bibinfo  {journal} {Nature}\ }\textbf {\bibinfo {volume} {323}},\ \bibinfo
  {pages} {533} (\bibinfo {year} {1986})}\BibitemShut {NoStop}%
\bibitem [{\citenamefont {{Werbos}}(1990)}]{bptt}%
  \BibitemOpen
  \bibfield  {author} {\bibinfo {author} {\bibfnamefont {P.~J.}\ \bibnamefont
  {{Werbos}}},\ }\href@noop {} {\bibfield  {journal} {\bibinfo  {journal}
  {Proceedings of the IEEE}\ }\textbf {\bibinfo {volume} {78}},\ \bibinfo
  {pages} {1550} (\bibinfo {year} {1990})}\BibitemShut {NoStop}%
\bibitem [{\citenamefont {Rivkind}\ and\ \citenamefont
  {Barak}(2017)}]{PhysRevLett.118.258101}%
  \BibitemOpen
  \bibfield  {author} {\bibinfo {author} {\bibfnamefont {A.}~\bibnamefont
  {Rivkind}}\ and\ \bibinfo {author} {\bibfnamefont {O.}~\bibnamefont
  {Barak}},\ }\href {\doibase 10.1103/PhysRevLett.118.258101} {\bibfield
  {journal} {\bibinfo  {journal} {Phys. Rev. Lett.}\ }\textbf {\bibinfo
  {volume} {118}},\ \bibinfo {pages} {258101} (\bibinfo {year}
  {2017})}\BibitemShut {NoStop}%
\bibitem [{Note1()}]{Note1}%
  \BibitemOpen
  \bibinfo {note} {Because $x_i$ takes both positive and negative values, it
  does not represent the neural activity per se. Instead, in the context of
  neuroscience, $(1+\protect \qopname \relax o{tanh}(x_i))/2$ is often regarded
  as the firing rate of the neuron.}\BibitemShut {Stop}%
\bibitem [{\citenamefont {Bishop}(2006)}]{PRML}%
  \BibitemOpen
  \bibfield  {author} {\bibinfo {author} {\bibfnamefont {C.~M.}\ \bibnamefont
  {Bishop}},\ }\href@noop {} {\emph {\bibinfo {title} {Pattern Recognition and
  Machine Learning}}}\ (\bibinfo  {publisher} {Springer},\ \bibinfo {year}
  {2006})\BibitemShut {NoStop}%
\bibitem [{Note2()}]{Note2}%
  \BibitemOpen
  \bibinfo {note} {The relationship between logit $z_1-z_2$ and the probability
  calculated by the softmax function is as follows. By definition of the
  softmax function, we have
  $p_1=e^{z_1}/(e^{z_1}+e^{z_2})=1/(1+e^{-(z_1-z_2)})$. Then, by the definition
  of logit function, we get the logit of $p_1$ as $\protect \qopname \relax
  o{log}(p_1/(1-p_1))=\protect \qopname \relax o{log}(e^{z_1-z_2})=z_1-z_2$.
  From the above explanation, the output of the softmax function can be treated
  as a probability because $p_1$ and $p_2$ satisfy $0\leq p_1,p_2 \leq 1$, and
  $p_1+p_2=e^{z_1}/(e^{z_1}+e^{z_2})+e^{z_2}/(e^{z_1}+e^{z_2})=1$. In this
  study, specifically, we interpret this probability as the probability that
  the first signal is recognized to have a higher frequency by the
  RNN.}\BibitemShut {Stop}%
\bibitem [{\citenamefont {Maass}\ \emph {et~al.}(2002)\citenamefont {Maass},
  \citenamefont {Natschläger},\ and\ \citenamefont {Markram}}]{echo-state}%
  \BibitemOpen
  \bibfield  {author} {\bibinfo {author} {\bibfnamefont {W.}~\bibnamefont
  {Maass}}, \bibinfo {author} {\bibfnamefont {T.}~\bibnamefont {Natschläger}},
  \ and\ \bibinfo {author} {\bibfnamefont {H.}~\bibnamefont {Markram}},\
  }\href@noop {} {\bibfield  {journal} {\bibinfo  {journal} {Neural Comput.}\
  }\textbf {\bibinfo {volume} {14(11)}} (\bibinfo {year} {2002})}\BibitemShut
  {NoStop}%
\bibitem [{\citenamefont {Sussillo}\ and\ \citenamefont
  {Abbott}(2009)}]{FORCE_learning}%
  \BibitemOpen
  \bibfield  {author} {\bibinfo {author} {\bibfnamefont {D.}~\bibnamefont
  {Sussillo}}\ and\ \bibinfo {author} {\bibfnamefont {L.}~\bibnamefont
  {Abbott}},\ }\href {\doibase https://doi.org/10.1016/j.neuron.2009.07.018}
  {\bibfield  {journal} {\bibinfo  {journal} {Neuron}\ }\textbf {\bibinfo
  {volume} {63}},\ \bibinfo {pages} {544} (\bibinfo {year} {2009})}\BibitemShut
  {NoStop}%
\bibitem [{\citenamefont {LeCun}\ \emph {et~al.}(2015)\citenamefont {LeCun},
  \citenamefont {Bengio},\ and\ \citenamefont {Hinton}}]{DeepLearning}%
  \BibitemOpen
  \bibfield  {author} {\bibinfo {author} {\bibfnamefont {Y.}~\bibnamefont
  {LeCun}}, \bibinfo {author} {\bibfnamefont {Y.}~\bibnamefont {Bengio}}, \
  and\ \bibinfo {author} {\bibfnamefont {G.}~\bibnamefont {Hinton}},\
  }\href@noop {} {\bibfield  {journal} {\bibinfo  {journal} {Nature}\ }\textbf
  {\bibinfo {volume} {521}},\ \bibinfo {pages} {436} (\bibinfo {year}
  {2015})}\BibitemShut {NoStop}%
\bibitem [{\citenamefont {Goodfellow}\ \emph {et~al.}(2016)\citenamefont
  {Goodfellow}, \citenamefont {Bengio},\ and\ \citenamefont
  {Courville}}]{deep_learning_book}%
  \BibitemOpen
  \bibfield  {author} {\bibinfo {author} {\bibfnamefont {I.}~\bibnamefont
  {Goodfellow}}, \bibinfo {author} {\bibfnamefont {Y.}~\bibnamefont {Bengio}},
  \ and\ \bibinfo {author} {\bibfnamefont {A.}~\bibnamefont {Courville}},\
  }\href@noop {} {\emph {\bibinfo {title} {Deep Learning}}}\ (\bibinfo
  {publisher} {MIT Press},\ \bibinfo {year} {2016})\ \bibinfo {note}
  {\url{http://www.deeplearningbook.org}}\BibitemShut {NoStop}%
\bibitem [{\citenamefont {Kingma}\ and\ \citenamefont {Ba}(2014)}]{adam}%
  \BibitemOpen
  \bibfield  {author} {\bibinfo {author} {\bibfnamefont {D.}~\bibnamefont
  {Kingma}}\ and\ \bibinfo {author} {\bibfnamefont {J.}~\bibnamefont {Ba}},\
  }\href@noop {} {\bibfield  {journal} {\bibinfo  {journal} {International
  Conference on Learning Representations}\ } (\bibinfo {year}
  {2014})}\BibitemShut {NoStop}%
\bibitem [{\citenamefont {Paszke}\ \emph {et~al.}(2019)\citenamefont {Paszke},
  \citenamefont {Gross},\ and\ \citenamefont {et~al}}]{NEURIPS2019_9015}%
  \BibitemOpen
  \bibfield  {author} {\bibinfo {author} {\bibfnamefont {A.}~\bibnamefont
  {Paszke}}, \bibinfo {author} {\bibfnamefont {S.}~\bibnamefont {Gross}}, \
  and\ \bibinfo {author} {\bibnamefont {et~al}},\ }in\ \href@noop {} {\emph
  {\bibinfo {booktitle} {Advances in Neural Information Processing Systems
  32}}}\ (\bibinfo  {publisher} {Neural Information Processing Systems
  Foundation},\ \bibinfo {year} {2019})\ pp.\ \bibinfo {pages}
  {8024--8035}\BibitemShut {NoStop}%
\bibitem [{\citenamefont {Lever}\ \emph {et~al.}(2017)\citenamefont {Lever},
  \citenamefont {Krzywinski},\ and\ \citenamefont {Altman}}]{PCA}%
  \BibitemOpen
  \bibfield  {author} {\bibinfo {author} {\bibfnamefont {J.}~\bibnamefont
  {Lever}}, \bibinfo {author} {\bibfnamefont {M.}~\bibnamefont {Krzywinski}}, \
  and\ \bibinfo {author} {\bibfnamefont {N.}~\bibnamefont {Altman}},\
  }\href@noop {} {\bibfield  {journal} {\bibinfo  {journal} {Nat Methods}\
  }\textbf {\bibinfo {volume} {14}},\ \bibinfo {pages} {641} (\bibinfo {year}
  {2017})}\BibitemShut {NoStop}%
\bibitem [{Note3()}]{Note3}%
  \BibitemOpen
  \bibinfo {note} {See Supplemental Material Fig.1 for the characteristics of
  the neural activities during the delay period.}\BibitemShut {Stop}%
\bibitem [{Note4()}]{Note4}%
  \BibitemOpen
  \bibinfo {note} {The fixed-point analysis\cite {barak_fixed_point} showed
  that there were a few slow points acting as pseudo-saddles through which the
  neural activity passed after the first signal. See Supplemental Material
  Fig.3 for the slow points plotted in the 3-dimensional PC space.}\BibitemShut
  {Stop}%
\bibitem [{Note5()}]{Note5}%
  \BibitemOpen
  \bibinfo {note} {See Supplemental Material Fig.2 for the neural activity of
  trained RNN with longer delay period.}\BibitemShut {Stop}%
\bibitem [{Note6()}]{Note6}%
  \BibitemOpen
  \bibinfo {note} {See Supplemental Material Fig.4 and 5 for the results of the
  different tasks.}\BibitemShut {Stop}%
\bibitem [{Note7()}]{Note7}%
  \BibitemOpen
  \bibinfo {note} {See Supplemental Material Fig.6 for the convergence time to
  the limit cycle.}\BibitemShut {Stop}%
\bibitem [{\citenamefont {Sirovich}(1989)}]{SIROVICH1989126}%
  \BibitemOpen
  \bibfield  {author} {\bibinfo {author} {\bibfnamefont {L.}~\bibnamefont
  {Sirovich}},\ }\href {\doibase https://doi.org/10.1016/0167-2789(89)90123-1}
  {\bibfield  {journal} {\bibinfo  {journal} {Physica D: Nonlinear Phenomena}\
  }\textbf {\bibinfo {volume} {37}},\ \bibinfo {pages} {126 } (\bibinfo {year}
  {1989})}\BibitemShut {NoStop}%
\bibitem [{\citenamefont {Ciliberto}\ and\ \citenamefont
  {Nicolaenko}(1991)}]{Ciliberto_1991}%
  \BibitemOpen
  \bibfield  {author} {\bibinfo {author} {\bibfnamefont {S.}~\bibnamefont
  {Ciliberto}}\ and\ \bibinfo {author} {\bibfnamefont {B.}~\bibnamefont
  {Nicolaenko}},\ }\href {\doibase 10.1209/0295-5075/14/4/003} {\bibfield
  {journal} {\bibinfo  {journal} {Europhysics Letters ({EPL})}\ }\textbf
  {\bibinfo {volume} {14}},\ \bibinfo {pages} {303} (\bibinfo {year}
  {1991})}\BibitemShut {NoStop}%
\bibitem [{Note8()}]{Note8}%
  \BibitemOpen
  \bibinfo {note} {See Supplemental Material Fig.7 for the estimated dimension
  of perturbed trajectories by the method in \cite {SIROVICH1989126,
  Ciliberto_1991}}\BibitemShut {NoStop}%
\bibitem [{\citenamefont {Crisanti}\ \emph {et~al.}(1988)\citenamefont
  {Crisanti}, \citenamefont {Paladin},\ and\ \citenamefont
  {Vulpiani}}]{lyapunov}%
  \BibitemOpen
  \bibfield  {author} {\bibinfo {author} {\bibfnamefont {A.}~\bibnamefont
  {Crisanti}}, \bibinfo {author} {\bibfnamefont {G.}~\bibnamefont {Paladin}}, \
  and\ \bibinfo {author} {\bibfnamefont {A.}~\bibnamefont {Vulpiani}},\
  }\href@noop {} {\bibfield  {journal} {\bibinfo  {journal} {J Stat Phys}\
  }\textbf {\bibinfo {volume} {53}},\ \bibinfo {pages} {583} (\bibinfo {year}
  {1988})}\BibitemShut {NoStop}%
\bibitem [{\citenamefont {Dale}(1935)}]{Dale}%
  \BibitemOpen
  \bibfield  {author} {\bibinfo {author} {\bibfnamefont {H.}~\bibnamefont
  {Dale}},\ }\href@noop {} {\bibfield  {journal} {\bibinfo  {journal} {Proc R
  Soc Med.}\ }\textbf {\bibinfo {volume} {28(3)}},\ \bibinfo {pages} {319}
  (\bibinfo {year} {1935})}\BibitemShut {NoStop}%
\bibitem [{\citenamefont {Song}\ \emph {et~al.}(2016)\citenamefont {Song},
  \citenamefont {Yang},\ and\ \citenamefont
  {Wang}}]{10.1371/journal.pcbi.1004792}%
  \BibitemOpen
  \bibfield  {author} {\bibinfo {author} {\bibfnamefont {H.~F.}\ \bibnamefont
  {Song}}, \bibinfo {author} {\bibfnamefont {G.~R.}\ \bibnamefont {Yang}}, \
  and\ \bibinfo {author} {\bibfnamefont {X.-J.}\ \bibnamefont {Wang}},\ }\href
  {\doibase 10.1371/journal.pcbi.1004792} {\bibfield  {journal} {\bibinfo
  {journal} {PLOS Computational Biology}\ }\textbf {\bibinfo {volume} {12}},\
  \bibinfo {pages} {1} (\bibinfo {year} {2016})}\BibitemShut {NoStop}%
\bibitem [{\citenamefont {Orhan}\ and\ \citenamefont {Ma}(2017)}]{Efficient}%
  \BibitemOpen
  \bibfield  {author} {\bibinfo {author} {\bibfnamefont {A.}~\bibnamefont
  {Orhan}}\ and\ \bibinfo {author} {\bibfnamefont {W.}~\bibnamefont {Ma}},\
  }\href@noop {} {\bibfield  {journal} {\bibinfo  {journal} {Nat Commun}\
  }\textbf {\bibinfo {volume} {8}} (\bibinfo {year} {2017})}\BibitemShut
  {NoStop}%
\bibitem [{\citenamefont {Bellec}\ \emph {et~al.}(2017)\citenamefont {Bellec},
  \citenamefont {Kappel}, \citenamefont {Maass},\ and\ \citenamefont
  {Legenstein}}]{deep-r}%
  \BibitemOpen
  \bibfield  {author} {\bibinfo {author} {\bibfnamefont {G.}~\bibnamefont
  {Bellec}}, \bibinfo {author} {\bibfnamefont {D.}~\bibnamefont {Kappel}},
  \bibinfo {author} {\bibfnamefont {W.}~\bibnamefont {Maass}}, \ and\ \bibinfo
  {author} {\bibfnamefont {R.~A.}\ \bibnamefont {Legenstein}},\ }\href
  {http://dblp.uni-trier.de/db/journals/corr/corr1711.html#abs-1711-05136}
  {\bibfield  {journal} {\bibinfo  {journal} {CoRR}\ }\textbf {\bibinfo
  {volume} {abs/1711.05136}} (\bibinfo {year} {2017})}\BibitemShut {NoStop}%
\bibitem [{Note9()}]{Note9}%
  \BibitemOpen
  \bibinfo {note} {See Supplemental Material Fig.8 and 9 for the results under
  the excitatory-inhibitory networks condition and the sparse networks
  condition}\BibitemShut {NoStop}%
\bibitem [{\citenamefont {Barak}(2017)}]{versatile}%
  \BibitemOpen
  \bibfield  {author} {\bibinfo {author} {\bibfnamefont {O.}~\bibnamefont
  {Barak}},\ }\href@noop {} {\bibfield  {journal} {\bibinfo  {journal} {Curr
  Opin Neurobiol.}\ }\textbf {\bibinfo {volume} {46}},\ \bibinfo {pages} {1}
  (\bibinfo {year} {2017})}\BibitemShut {NoStop}%
\bibitem [{\citenamefont {Richards}\ \emph {et~al.}(2019)\citenamefont
  {Richards}, \citenamefont {Lillicrap}, \citenamefont {Beaudoin},\ and\
  \citenamefont {et~al.}}]{Richards2019}%
  \BibitemOpen
  \bibfield  {author} {\bibinfo {author} {\bibfnamefont {B.}~\bibnamefont
  {Richards}}, \bibinfo {author} {\bibfnamefont {T.}~\bibnamefont {Lillicrap}},
  \bibinfo {author} {\bibfnamefont {P.}~\bibnamefont {Beaudoin}}, \ and\
  \bibinfo {author} {\bibnamefont {et~al.}},\ }\href@noop {} {\bibfield
  {journal} {\bibinfo  {journal} {Nat Neurosci}\ }\textbf {\bibinfo {volume}
  {22}},\ \bibinfo {pages} {1761} (\bibinfo {year} {2019})}\BibitemShut
  {NoStop}%
\bibitem [{\citenamefont {Christian}\ \emph {et~al.}(2005)\citenamefont
  {Christian}, \citenamefont {Ranulfo},\ and\ \citenamefont
  {Carlos}}]{romo_model}%
  \BibitemOpen
  \bibfield  {author} {\bibinfo {author} {\bibfnamefont {K.~M.}\ \bibnamefont
  {Christian}}, \bibinfo {author} {\bibfnamefont {R.}~\bibnamefont {Ranulfo}},
  \ and\ \bibinfo {author} {\bibfnamefont {D.~B.}\ \bibnamefont {Carlos}},\
  }\href@noop {} {\bibfield  {journal} {\bibinfo  {journal} {science}\ }\textbf
  {\bibinfo {volume} {307}},\ \bibinfo {pages} {1121} (\bibinfo {year}
  {2005})}\BibitemShut {NoStop}%
\bibitem [{\citenamefont {Seung}(1996)}]{Seung13339}%
  \BibitemOpen
  \bibfield  {author} {\bibinfo {author} {\bibfnamefont {H.~S.}\ \bibnamefont
  {Seung}},\ }\href {\doibase 10.1073/pnas.93.23.13339} {\bibfield  {journal}
  {\bibinfo  {journal} {Proceedings of the National Academy of Sciences}\
  }\textbf {\bibinfo {volume} {93}},\ \bibinfo {pages} {13339} (\bibinfo {year}
  {1996})}\BibitemShut {NoStop}%
\bibitem [{\citenamefont {Mante}\ and\ \citenamefont
  {et~al.}(2013)}]{Mante2013}%
  \BibitemOpen
  \bibfield  {author} {\bibinfo {author} {\bibfnamefont {V.}~\bibnamefont
  {Mante}}\ and\ \bibinfo {author} {\bibnamefont {et~al.}},\ }\href {\doibase
  10.1038/nature12742} {\bibfield  {journal} {\bibinfo  {journal} {Nature}\
  }\textbf {\bibinfo {volume} {503}},\ \bibinfo {pages} {78} (\bibinfo {year}
  {2013})}\BibitemShut {NoStop}%
\bibitem [{Note10()}]{Note10}%
  \BibitemOpen
  \bibinfo {note} {For the possible applicability of chaotic attractors to
  short-term memory, see J. S. Nicolis and I. Tsuda, Chaotic dynamics of
  information processing: The ``magic number seven plus-minus two’’
  revisited, Bulletin of Mathematical Biology47, 343(1985).}\BibitemShut
  {Stop}%
\bibitem [{\citenamefont {Mongillo}\ \emph {et~al.}(2008)\citenamefont
  {Mongillo}, \citenamefont {Barak},\ and\ \citenamefont
  {Tsodyks}}]{Mongillo1543}%
  \BibitemOpen
  \bibfield  {author} {\bibinfo {author} {\bibfnamefont {G.}~\bibnamefont
  {Mongillo}}, \bibinfo {author} {\bibfnamefont {O.}~\bibnamefont {Barak}}, \
  and\ \bibinfo {author} {\bibfnamefont {M.}~\bibnamefont {Tsodyks}},\ }\href
  {\doibase 10.1126/science.1150769} {\bibfield  {journal} {\bibinfo  {journal}
  {Science}\ }\textbf {\bibinfo {volume} {319}},\ \bibinfo {pages} {1543}
  (\bibinfo {year} {2008})}\BibitemShut {NoStop}%
\bibitem [{\citenamefont {Mi}\ \emph {et~al.}(2017)\citenamefont {Mi},
  \citenamefont {Katkov},\ and\ \citenamefont {Tsodyks}}]{Tsodyks2017}%
  \BibitemOpen
  \bibfield  {author} {\bibinfo {author} {\bibfnamefont {Y.}~\bibnamefont
  {Mi}}, \bibinfo {author} {\bibfnamefont {M.}~\bibnamefont {Katkov}}, \ and\
  \bibinfo {author} {\bibfnamefont {M.}~\bibnamefont {Tsodyks}},\ }\href
  {\doibase 10.1016/j.neuron.2016.12.004} {\bibfield  {journal} {\bibinfo
  {journal} {Neuron}\ }\textbf {\bibinfo {volume} {93}},\ \bibinfo {pages}
  {323—330} (\bibinfo {year} {2017})}\BibitemShut {NoStop}%
\bibitem [{\citenamefont {Taher}\ \emph {et~al.}(2020)\citenamefont {Taher},
  \citenamefont {Torcini},\ and\ \citenamefont {Olmi}}]{10.1371}%
  \BibitemOpen
  \bibfield  {author} {\bibinfo {author} {\bibfnamefont {H.}~\bibnamefont
  {Taher}}, \bibinfo {author} {\bibfnamefont {A.}~\bibnamefont {Torcini}}, \
  and\ \bibinfo {author} {\bibfnamefont {S.}~\bibnamefont {Olmi}},\ }\href
  {\doibase 10.1371/journal.pcbi.1008533} {\bibfield  {journal} {\bibinfo
  {journal} {PLOS Computational Biology}\ }\textbf {\bibinfo {volume} {16}},\
  \bibinfo {pages} {1} (\bibinfo {year} {2020})}\BibitemShut {NoStop}%
\bibitem [{\citenamefont {Muller}\ \emph {et~al.}(2018)\citenamefont {Muller},
  \citenamefont {Chavane}, \citenamefont {Reynolds},\ and\ \citenamefont
  {et~al.}}]{travellingwave}%
  \BibitemOpen
  \bibfield  {author} {\bibinfo {author} {\bibfnamefont {L.}~\bibnamefont
  {Muller}}, \bibinfo {author} {\bibfnamefont {F.}~\bibnamefont {Chavane}},
  \bibinfo {author} {\bibfnamefont {J.}~\bibnamefont {Reynolds}}, \ and\
  \bibinfo {author} {\bibnamefont {et~al.}},\ }\href@noop {} {\bibfield
  {journal} {\bibinfo  {journal} {Nat Rev Neurosci}\ }\textbf {\bibinfo
  {volume} {19}},\ \bibinfo {pages} {255} (\bibinfo {year} {2018})}\BibitemShut
  {NoStop}%
\bibitem [{\citenamefont {H.}\ \emph {et~al.}(2018)\citenamefont {H.},
  \citenamefont {AJ.},\ and\ \citenamefont {et~al.}}]{travellingwave2}%
  \BibitemOpen
  \bibfield  {author} {\bibinfo {author} {\bibfnamefont {Z.}~\bibnamefont
  {H.}}, \bibinfo {author} {\bibfnamefont {W.}~\bibnamefont {AJ.}}, \ and\
  \bibinfo {author} {\bibnamefont {et~al.}},\ }\href@noop {} {\bibfield
  {journal} {\bibinfo  {journal} {Neuron}\ }\textbf {\bibinfo {volume} {27}},\
  \bibinfo {pages} {1269} (\bibinfo {year} {2018})}\BibitemShut {NoStop}%
\bibitem [{\citenamefont {Warasinee}\ \emph {et~al.}(2017)\citenamefont
  {Warasinee}, \citenamefont {Xiao-Jing},\ and\ \citenamefont
  {et~al.}}]{transient_robustness}%
  \BibitemOpen
  \bibfield  {author} {\bibinfo {author} {\bibfnamefont {C.}~\bibnamefont
  {Warasinee}}, \bibinfo {author} {\bibfnamefont {W.}~\bibnamefont
  {Xiao-Jing}}, \ and\ \bibinfo {author} {\bibnamefont {et~al.}},\ }\href@noop
  {} {\bibfield  {journal} {\bibinfo  {journal} {Neuron}\ }\textbf {\bibinfo
  {volume} {93}},\ \bibinfo {pages} {1504} (\bibinfo {year}
  {2017})}\BibitemShut {NoStop}%
\bibitem [{\citenamefont {Waddington}(1957)}]{Waddington}%
  \BibitemOpen
  \bibfield  {author} {\bibinfo {author} {\bibfnamefont {C.~H.}\ \bibnamefont
  {Waddington}},\ }\href@noop {} {\emph {\bibinfo {title} {The strategy of the
  genes: A discussion of some aspects of theoretical biology}}}\ (\bibinfo
  {publisher} {Allen \& Unwin},\ \bibinfo {address} {London},\ \bibinfo {year}
  {1957})\BibitemShut {NoStop}%
\bibitem [{\citenamefont {Matsushita}\ and\ \citenamefont
  {Kaneko}(2020)}]{Matsushita}%
  \BibitemOpen
  \bibfield  {author} {\bibinfo {author} {\bibfnamefont {Y.}~\bibnamefont
  {Matsushita}}\ and\ \bibinfo {author} {\bibfnamefont {K.}~\bibnamefont
  {Kaneko}},\ }\href {\doibase 10.1103/PhysRevResearch.2.023083} {\bibfield
  {journal} {\bibinfo  {journal} {Phys. Rev. Research}\ }\textbf {\bibinfo
  {volume} {2}},\ \bibinfo {pages} {023083} (\bibinfo {year}
  {2020})}\BibitemShut {NoStop}%
\bibitem [{\citenamefont {Young}\ \emph {et~al.}(2017)\citenamefont {Young},
  \citenamefont {Hatakeyama},\ and\ \citenamefont {Kaneko}}]{Jon}%
  \BibitemOpen
  \bibfield  {author} {\bibinfo {author} {\bibfnamefont {J.~T.}\ \bibnamefont
  {Young}}, \bibinfo {author} {\bibfnamefont {T.~S.}\ \bibnamefont
  {Hatakeyama}}, \ and\ \bibinfo {author} {\bibfnamefont {K.}~\bibnamefont
  {Kaneko}},\ }\href {\doibase 10.1371/journal.pcbi.1005434} {\bibfield
  {journal} {\bibinfo  {journal} {PLOS Computational Biology}\ }\textbf
  {\bibinfo {volume} {13}},\ \bibinfo {pages} {1} (\bibinfo {year}
  {2017})}\BibitemShut {NoStop}%
\end{thebibliography}%

\clearpage
\begin{widetext}
\begin{center}
\textbf{\LARGE Supplemental Material}
\end{center}
\end{widetext}

\setcounter{figure}{0}
\renewcommand{\thefigure}{S-\arabic{figure}}
\widetext

\begin{figure}[H]
    \center
    \includegraphics[width=8cm]{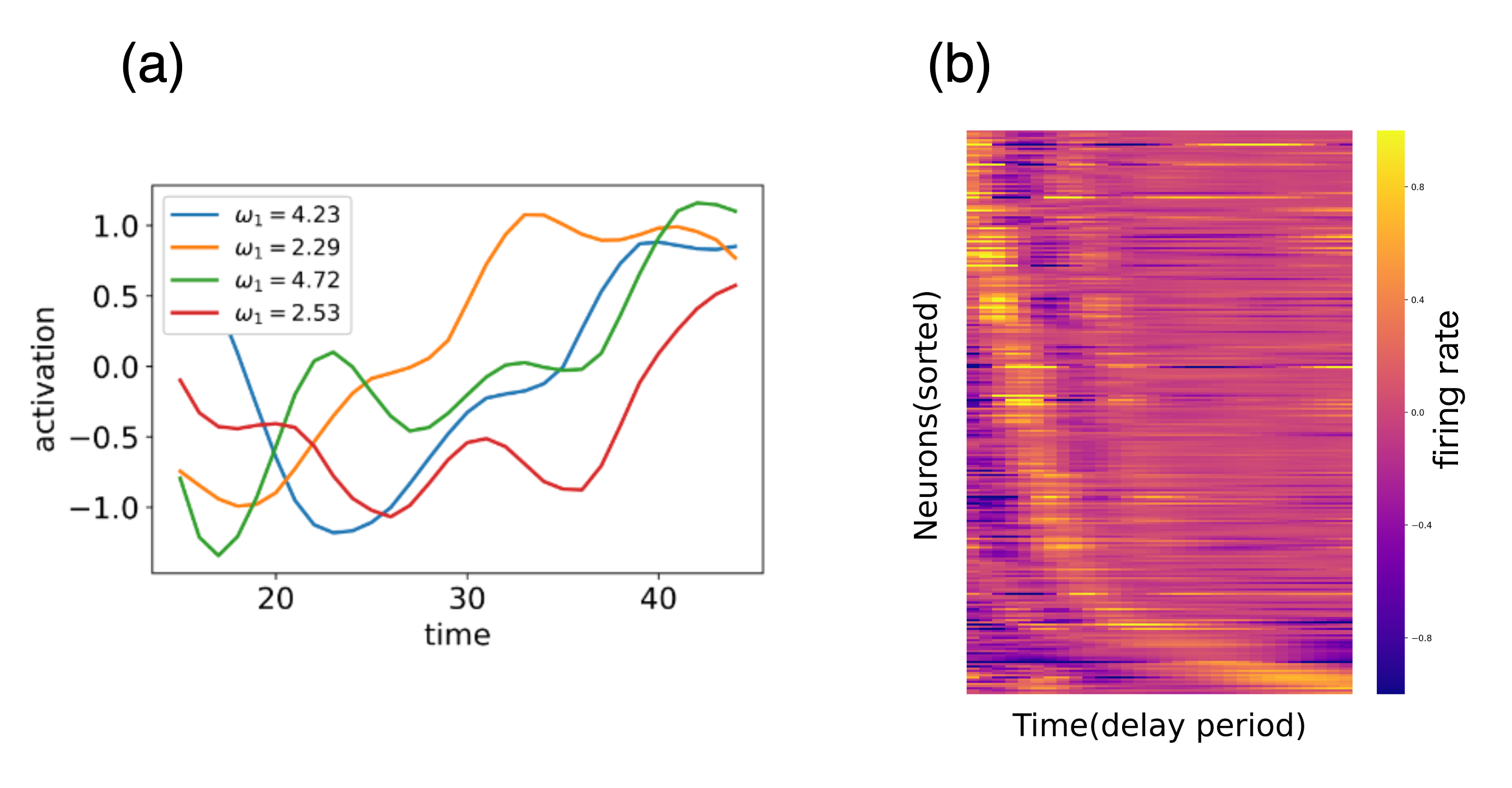}
    \caption{(a)The neural activity of one neuron $x_{100}(t)$ during the delay period. Different colors represent the activity for different $\omega_1$. (b) The firing rate of neurons during the delay period(Raster plot). Neurons are sorted by the time of their peak firing. One can see the propagation of neural activities, whose 'speed' slows down during the delay period. As shown in the figure, the neural activity changes dynamically during the delay period. This means that stable short-term memories are encoded by transient neural dynamics. The dynamics slow down during the delay period so that the short-term memory is maintained for a sufficient time. Note that although neural activities slow down, they converge not to a fixed point, but to the limit cycle. 
    }
    \label{fig:dynamical_memory}
\end{figure}

\begin{figure*}
    \center
    \includegraphics[width=15cm]{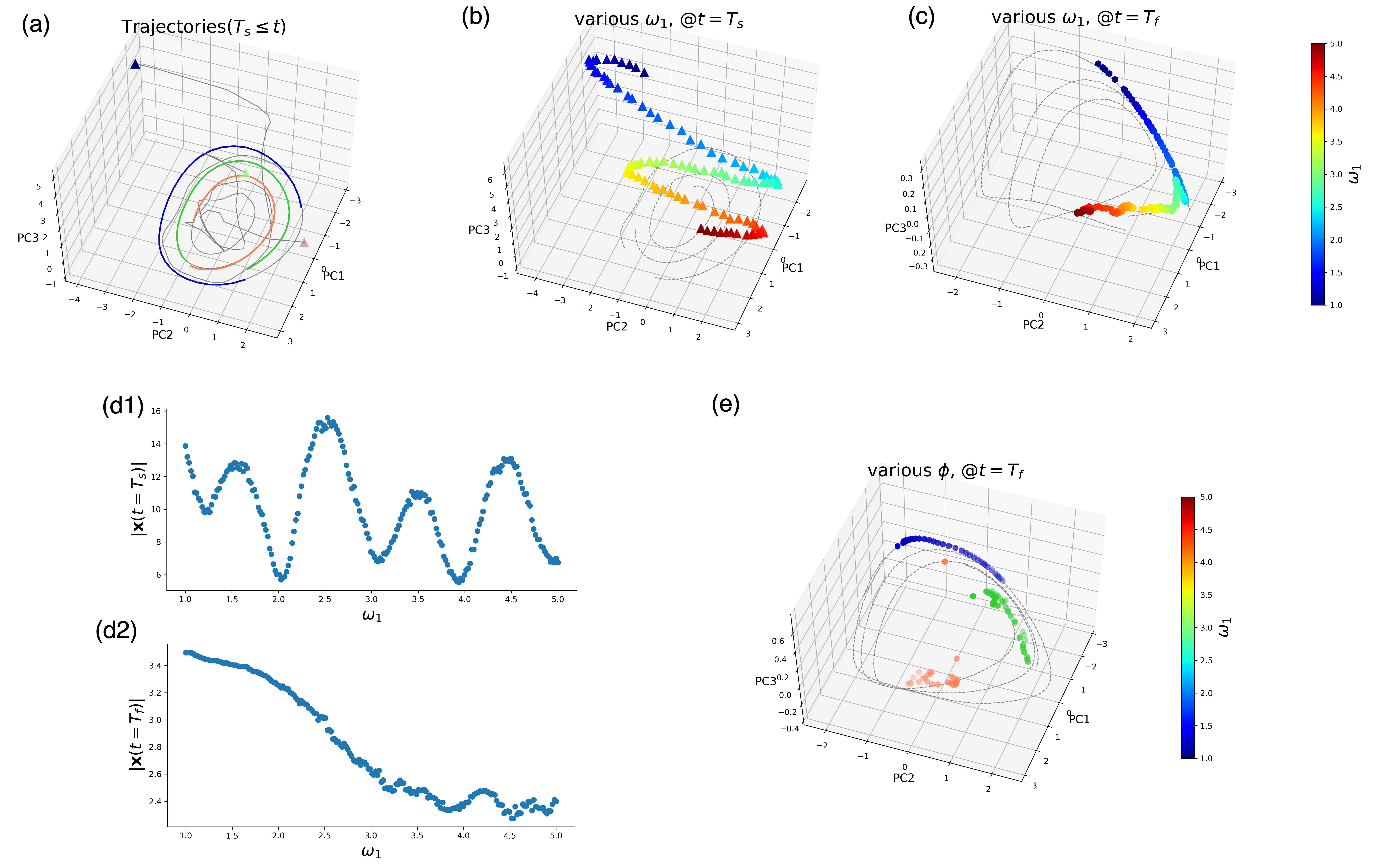}
    \caption{The neural activity of trained RNN with the condition that in the training phase, $T_d$ is homogeneously distributed as $T_d \in [75,105]$ and in the test phase, $T_d$ is fixed at 90.(c.f. in main paper, in the training phase, $T_d$ is homogeneously distributed as $T_d \in [25,35]$ and in the test phase, $T_d$ is fixed at 30). \\
    (a)Trajectories of neural activities during prolonged delay period($T_s \leq t$, without second signal), were plotted in 3-dimensional PC space. Trajectories from three different $\omega_1$, giving rise to different states at $t=T_s$, were given by triangle symbol.(b) Neural states $\bf x$ in the PC space at the beginning of the delay period($t=T_s$) for 50 different $\omega_1$'s presented in different colors.  (c)Neural states $\bf x$ in the PC space at the end of the delay period($t=T_f$) for $\omega_1$ corresponding to (b). (d1) Scatter plot of the norm $|{\bf x}(t=T_s)|$ against the frequency of the first signal $\omega_1$.(d2) Scatter plot of the norm at the end of the delay period $|{\bf x}(t=T_f)|$ against the frequency of the first signal $\omega_1$. Monotonic dependence is discernible. (e)Neural states $\bf x$ in the PC space at the end of the delay period($t=T_f$) for $\omega_1=1.5, 3, 4.5$. with various $0\leq \phi \leq \pi$. \\
    As in the case of $T_d=30$(main paper, Fig.3), a transient oscillation is formed during the delay period, and at $t=T_f$, $\omega_1$ is encoded as amplitude of transient oscillation ($|{\bf x}(t=T_f)|$ has a monotonic dependence on $\omega_1$).
    }
    \label{fig:dynamical_memory}
\end{figure*}

\begin{figure*}
    \center
    \includegraphics[width=12cm]{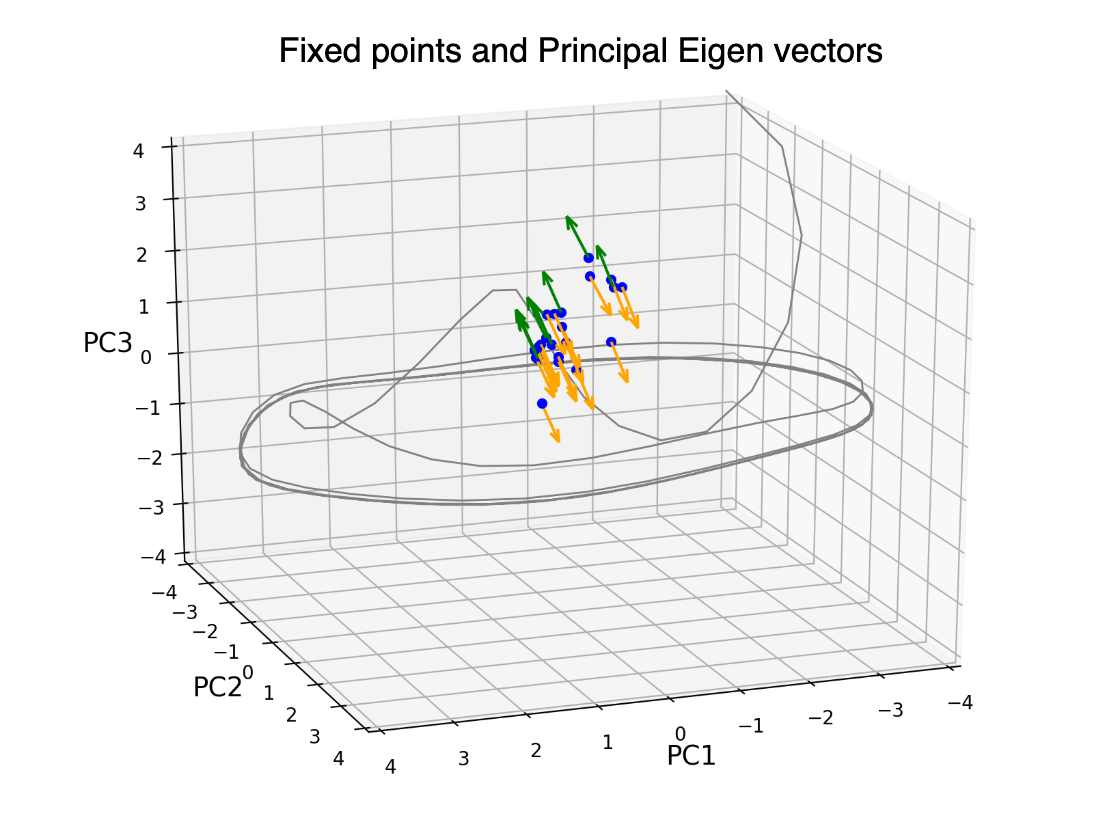}
    \caption{'Slow points' plotted in the 3-dimensional PC space with eigenvectors corresponding to unstable modes. The speed at a point on the trajectory $|{\bf \dot x}|_2 = |-{\bf x}+{\bf J}\tanh({\bf x})|_2$ was computed, and its local minimum points were obtained using the gradient method. The fixed points where the speed is completely zero are only the central points of the limit cycle, but there are some 'slow points' where the speed is very less, as depicted in the figure. These 'slow points' are also useful to understand the nature of the dynamics because they often work as pseudo stable plane or pseudo saddles; the eigenvalues of the Jacobian around these 'slow points' were calculated. It was found that the trajectory is attracted to these 'slow points', because most of the modes are negative(only one mode is positive, so they work as pseudo saddles). Indeed, the timing of neural activity slows down(Fig.\ref{fig:dynamical_memory}b) coincides with the timing of passage through these 'slow points', indicating that they contribute to the formation of the slow manifold.
    }
\end{figure*}

\begin{figure*}
    \center
    \includegraphics[width=15cm]{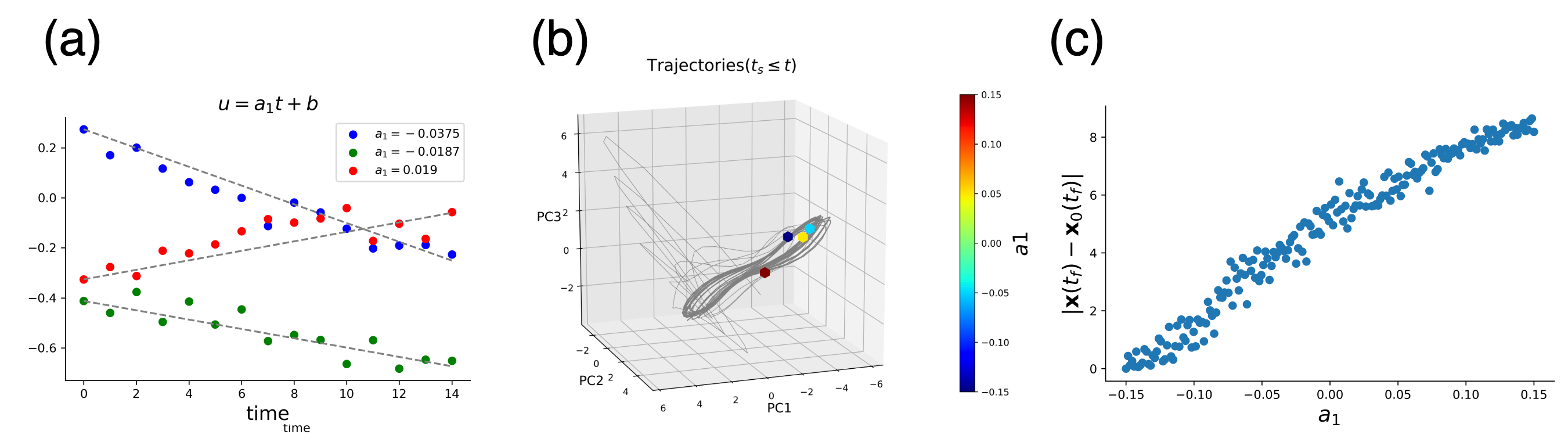}
    \caption{\label{fig:velocity_comparison} RNN was trained for velocity comparison task in which the first and second signals are given by $u_{1,2}(t)=a_{1,2}t+b+\eta$. Then, it was asked to determine which of the velocities $a_{1,2}$ is larger. $\eta$ is the Gaussian white noise.\\ (a)Sample signals. Different colors show different values of $a_1$. (b)Neural activity of trained RNN plotted in 3-dimensional PC space. Color points show the neural state at the end of the delay period ($t=T_f$). Oscillatory neural activity, which converged to a limit cycle after a long period of time, was observed during the delay period, (c) Scatter plot of the norm at the end of the delay period relative to the base state $|{\bf x}-{\bf x}_0|$ plotted against the frequency of the first signal $a_1$. The base state corresponds to the neural activity for the input signal's velocity which is $a_1=-0.15$. The input signal information is embedded in the amplitude of the transient oscillation.
}
\end{figure*}

\begin{figure*}
    \center
    \includegraphics[width=15cm]{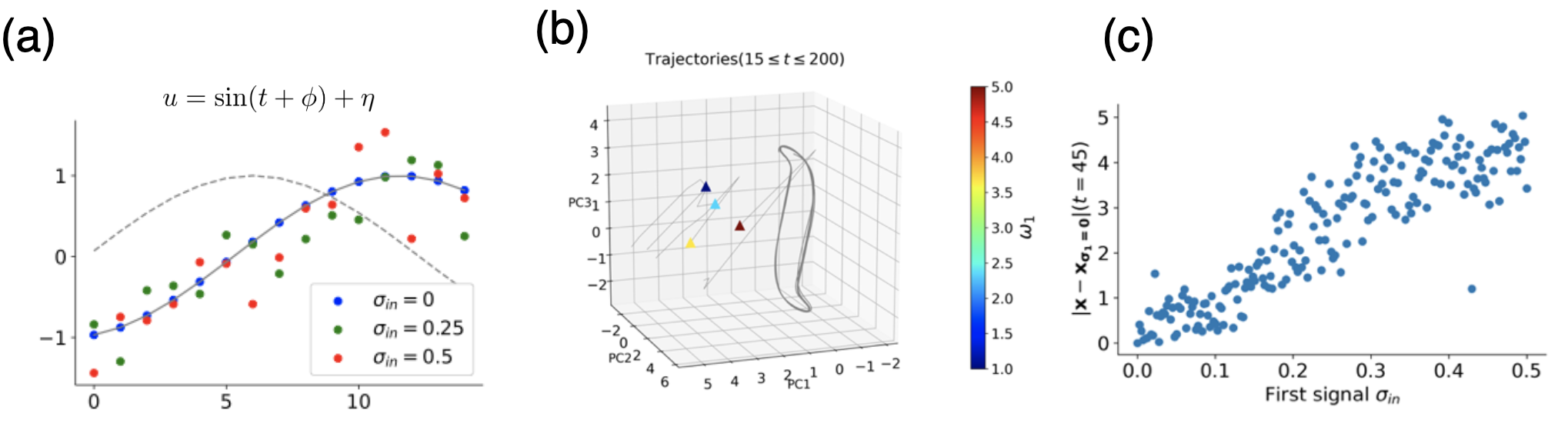}
    \caption{RNN was trained for noise variance comparison task. Here, the first and second signals are given by $u_{1,2}=\sin(t+\phi )+\eta$ in which $\eta$ as random Gaussian variable following $\eta \sim \mathcal N (0,\  \sigma_{1,2}^{in})$.Then, it is asked which of the variance $\sigma_{1,2}$ is larger. \\
    (a)Sample signals. Different colors show different $\sigma_1^{in}$. (b)Neural activity of trained RNN plotted in 3-dimensional PC space. Color points show the neural state at the end of the delay period ($t=T_f$). Oscillatory neural activity was observed during the delay period, which converged to a limit cycle after a long period of time. (c) Scatter plot of the norm at the end of the delay period relative to the base state $|{\bf x}-{\bf x}_0|$ plotted against the frequency of the first signal $a_1$. The base state corresponds to the neural activity for the input signal velocity with $\sigma_{in}=0$. The input signal information is embedded in the amplitude of the transient oscillation.}
\end{figure*}

\begin{figure*}
    \center
    \includegraphics[width=15cm]{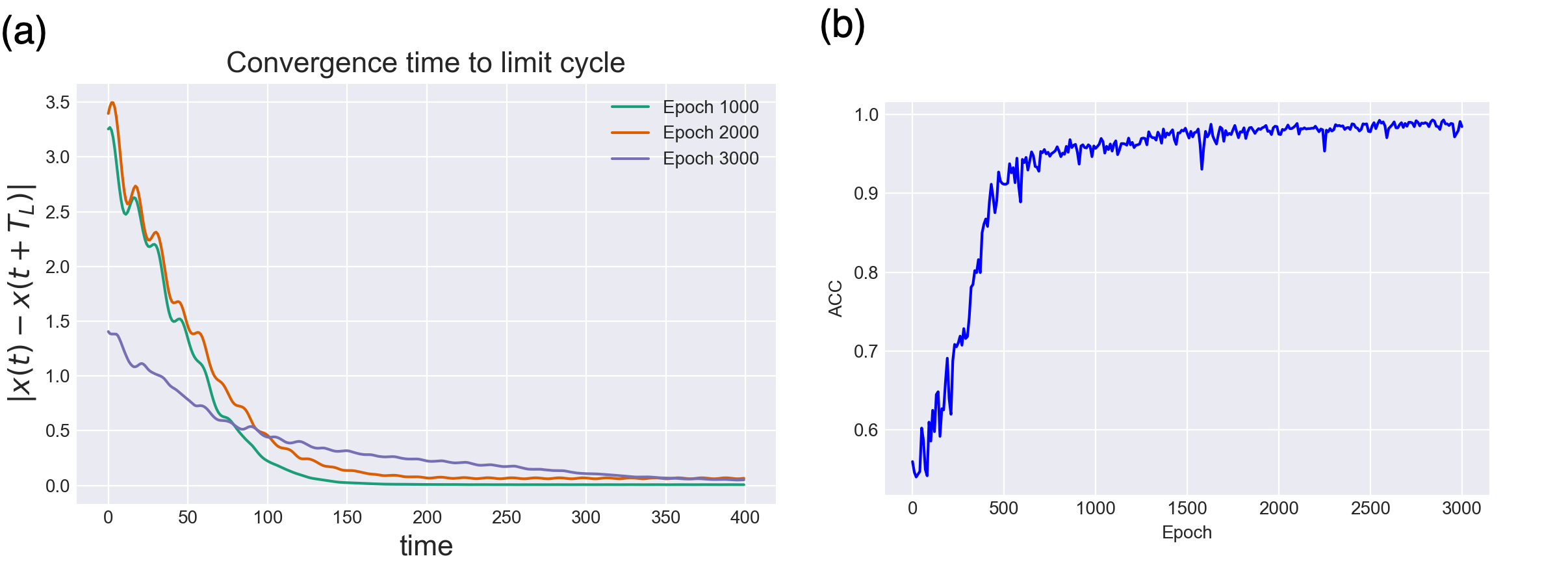}
    \caption{We estimated the convergence time to the limit cycle as the time $t$ when $|{\bf x}(t)-{\bf x}(t+T_L)|$ was sufficiently small. \\
    (a)$|{\bf x}(t)-{\bf x}(t+T_L)|$ with the time from $t=T_s$. Each of the three lines corresponds to the results of the RNN model under training (Epoch=1000, 2000, 3000). Since the delay period is 30, the convergence time is sufficiently larger than the delay period in all cases. In addition, the convergence  time increases as training progresses. (b) Time course of learning. It reaches sufficient accuracy before Epoch 1000, and from there on, the accuracy slowly improves as learning progresses.
    }
    \label{fig:dynamical_memory}
\end{figure*}

\begin{figure*}
    \center
    \includegraphics[width=12cm]{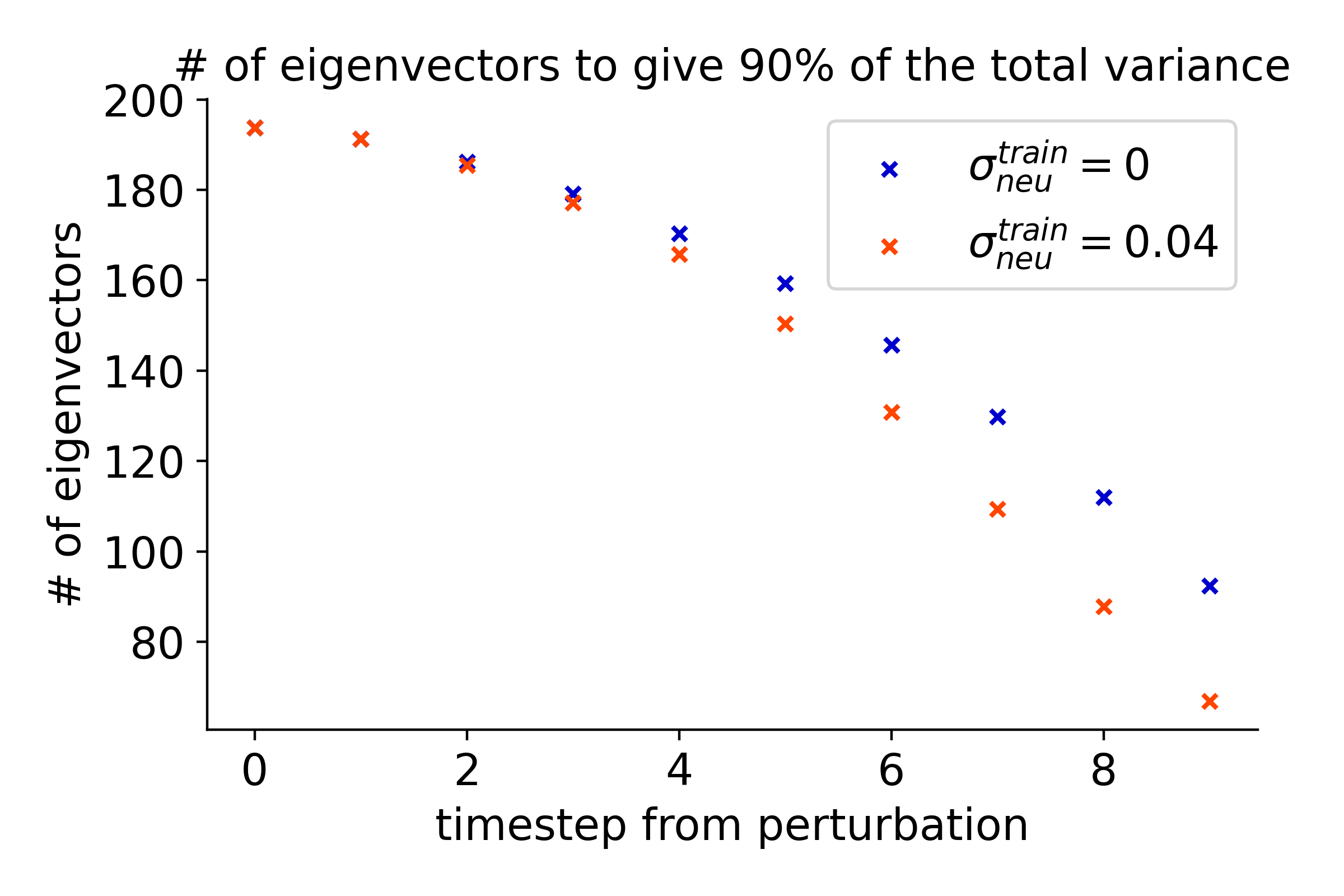}
    \caption{We estimated dimension of perturbed trajectories by the method in [35, 36], i.e. the number of eigenvectors of principal components need to give 90\% of the total variance. This value is smaller when the perturbed trajectory is restricted to a lower-dimensional manifold. The number of eigenvectors is plotted against the time after the application of perturbation, for $\sigma_{\rm neu}^{\rm train}=0$(blue), and $\sigma_{\rm neu}^{\rm train}=0.04$(red). As in the result of main paper Fig.4., the dynamics after the perturbation are more restricted to a lower-dimensional manifold for the model trained with noise.
    }
    \label{fig:dynamical_memory}
\end{figure*}

\begin{figure*}
    \center
    \includegraphics[width=15cm]{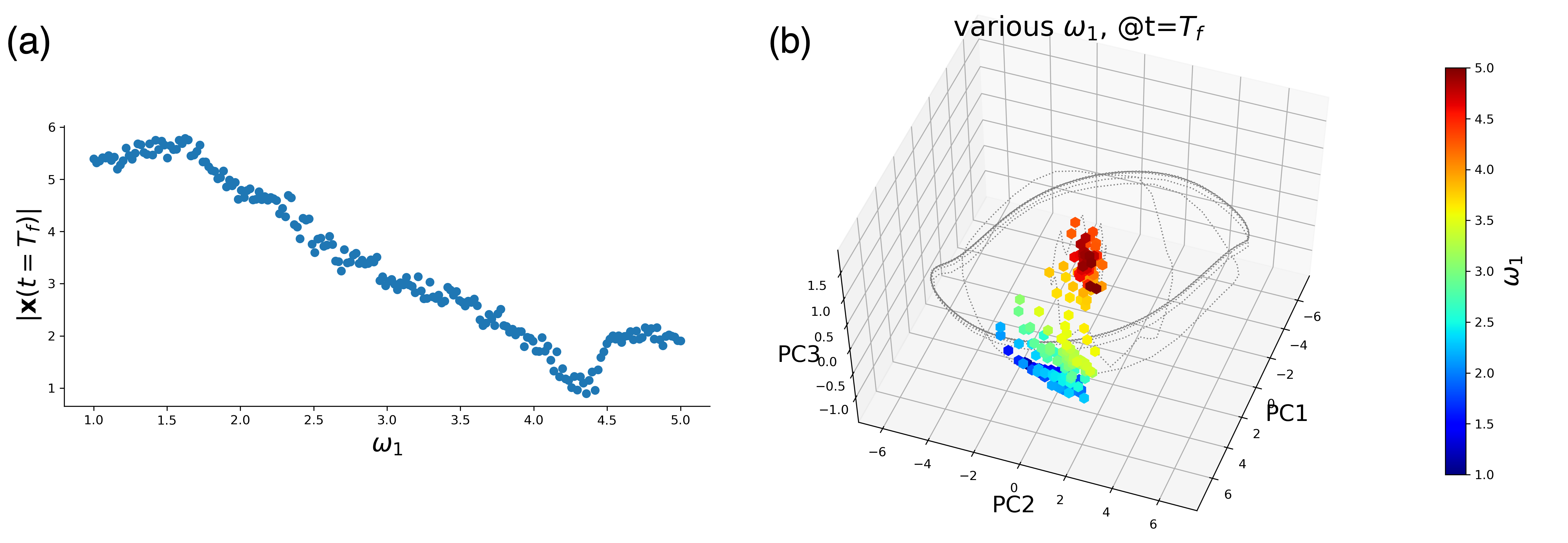}
    \caption{We trained RNN to solve the frequency comparison task while satisfying E-I networks condition. In this condition, hidden neurons were divided into excitatory-inhibitory neurons in a 4-to-1 ratio. Excitatory neurons only have 'excitatory synapes' which has positive value of synaptic weight, and Inhibitory neurons only have 'inhibitory synapes' which has negative value of synaptic weight. As in the result of non constrained RNN(main paper), there are transient oscillation and amplitude coding of input signals. \\
    (a) Scatter plot of the norm at the end of the delay period $|{\bf x}(t=T_f)|$ against the frequency of the first signal $\omega_1$. Monotonic dependence is discernible. (b) Neural states $\bf x$ in the PC space at the end of the delay period($t=T_f$) for different $\omega_1$.
 }
\end{figure*}

\begin{figure*}
    \center
    \includegraphics[width=15cm]{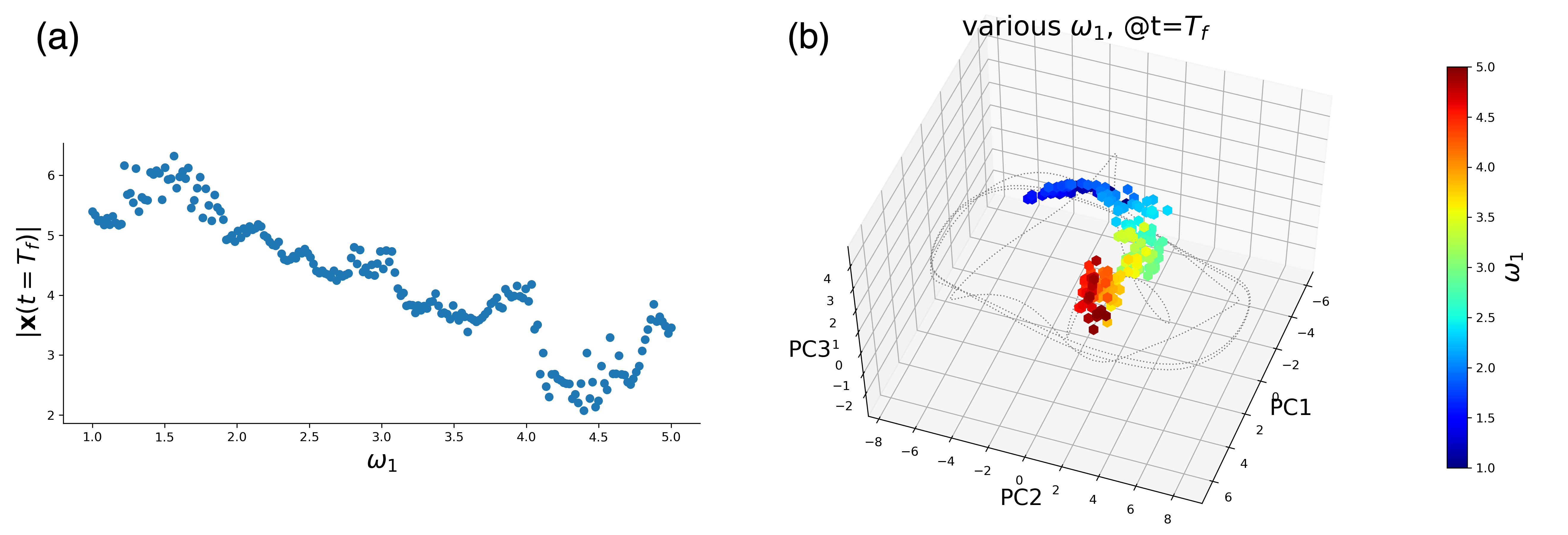}
    \caption{We trained RNN to solve the frequency comparison task while satisfying Sparse networks condition. In this condition, the percentage of synapses with non-zero weight is set to 20\%. As in the result of non constrained RNN(main paper), there are transient oscillation and amplitude coding of input signals. \\
    (a) Scatter plot of the norm at the end of the delay period $|{\bf x}(t=T_f)|$ against the frequency of the first signal $\omega_1$. Monotonic dependence is discernible. (b) Neural states $\bf x$ in the PC space at the end of the delay period($t=T_f$) for different $\omega_1$.
}
\end{figure*}

\end{document}